\documentclass[journal, font=10pt]{IEEEtran}

\usepackage{subcaption}
\usepackage{amsmath,amssymb,amsfonts}
\usepackage{algorithmic}
\usepackage{graphicx}
\usepackage{textcomp}
\usepackage{enumitem}
\usepackage{xcolor}
\usepackage{soul}
\usepackage{xkeyval}
\def\BibTeX{{\rm B\kern-.05em{\sc i\kern-.025em b}\kern-.08em
    T\kern-.1667em\lower.7ex\hbox{E}\kern-.125emX}}

\providecommand{\keywords}[1]{\textbf{\textit{Index terms---}} #1}
\usepackage[
  backend=biber,
  style=ieee,
  doi=true,
  url=false,
  isbn=false,
  eprint=false 
]{biblatex}
\AtBeginBibliography{\small}

\begin{document}

\title{Multi-photon QKD for Practical Quantum Networks}

\author{%
  \begin{minipage}[t]{0.32\textwidth}\centering
    \IEEEauthorblockN{Nitin Jha}\\
    \IEEEauthorblockA{%
      Department of Computer Science\\
      Kennesaw State University\\
      njha1@students.kennesaw.edu}
  \end{minipage}\hfill
  \begin{minipage}[t]{0.32\textwidth}\centering
    \IEEEauthorblockN{Abhishek Parakh}\\
    \IEEEauthorblockA{%
      Department of Computer Science\\
      Kennesaw State University\\
      aparakh@kennesaw.edu}
  \end{minipage}\hfill
  \begin{minipage}[t]{0.32\textwidth}\centering
    \IEEEauthorblockN{Mahadevan Subramaniam}\\
    \IEEEauthorblockA{%
      Computer Science Department\\
      University of Nebraska Omaha\\
      msubramaniam@unomaha.edu}
  \end{minipage}
}

\maketitle
\bstctlcite{IEEEexample:BSTcontrol}
\begin{abstract}
Quantum key distribution (QKD) will most likely be an integral part of any practical quantum network in the future. However, not all QKD protocols can be used in today's networks because of the lack of single-photon emitters and noisy intermediate quantum hardware. Attenuated-photon transmission, typically used to simulate single-photon emitters, severely limits the achievable transmission distances and makes the integration of the QKD into existing classical networks, that use tens of thousands of photons per bit of transmission, difficult. Furthermore, it has been found that protocol performance varies with topology. In order to remove the reliance of QKD on single-photon emitters and increase transmission distances, it is worthwhile to explore QKD protocols that do not rely on single-photon transmissions for security, such as the 3-stage QKD protocol, which can tolerate multiple photons in each burst without information leakage. This paper compares and contrasts the 3-stage QKD protocol with conventional QKD protocols and its efficiency in different network topologies and conditions. Furthermore, we establish a mathematical relationship between achievable key rates to increase transmission distances in various topologies. 
\end{abstract}


\section{Introduction}
\label{sec:intro}  

Quantum information theory has been of significant interest over recent decades, leading to advances in computing, networking, and sensing technologies \cite{Zhang2021DistributedQS, Aslam2023QuantumSF, Zhang2022FutureQC}. Quantum  networks, under specific hardware configurations, offer enhanced security for communications \cite{renner2008} and are seen as a vital element in the future development of the quantum internet. Despite some challenges \cite{NSA}, Quantum Key Distribution (QKD) remains a foundational technique for ensuring network security. Research efforts are currently directed towards developing near-ideal quantum hardware, perfecting QKD systems, and mitigating potential eavesdropping risks \cite{elliott2002building, yurke1984quantum, satoh2021attacking}. Practical implementations have been demonstrated through the DARPA network and other networks across Europe \cite{peev2009secoqc} and Tokyo \cite{sasaki2011field}, showing the feasibility of quantum networks. Recent developments include the deployment of a private quantum network across Europe during the 2021 G20 Summit in Trieste, featuring two sender and two receiver nodes \cite{ribezzo2023deploying}. Furthermore, from 2003 to 2016, the Chinese Academy of Sciences (CAS) developed a satellite and ground-based network, achieving secure key distribution over $7,600$ km using a decoy-state QKD transmitter onboard a low-earth-orbit satellite, which also facilitated a secure video conferencing session \cite{SatelliteToGroundQKD2017}. These advancements underscore the rapid progress in quantum technologies that are poised to support large-scale secure quantum communications \cite{wehner2018quantum, satoh2021attacking}.

QKD offers unconditional security because it utilizes the principles of quantum physics, making it \textit{future proof} against eavesdropping, unlike classical transmission methods \cite{unruh2013advances, gaidash2016revealing, bennett1992quantum, bennett2014quantum, ekert1991quantum}. QKD, combined with \textit{quantum-resistant classical algorithms} and \textit{quantum cryptography}, is expected to play a crucial role in securing future communications for applications such as smart grids and defense networks \cite{diamantinature}. Despite its potential, QKD faces several challenges, both theoretical and practical. These include photon number splitting (PNS) attacks, hardware inconsistencies \cite{Sajeed2021, verma2019photonnumber}, and vulnerabilities related to Bell inequality test \textit{loopholes} \cite{hensen2015}. Furthermore, practical issues with optical switches and trusted nodes limit network robustness and range, with compromised nodes posing significant security risks \cite{chen2021implementation}. Current approaches aim to manage multi-photon bursts, but these methods are limited because they still discard multi-photon bursts and rely on single photons, making integration with classical networks challenging. An alternative is the use of advanced multiphoton QKD protocols, such as the 3-stage protocol, which do not depend on single photons and accommodate higher photon burst rates \cite{Kak2006-3Stage, jha2024, parakh2018bootstrapped}. Recent studies have also shown potential for using the three-stage protocol in quantum secure direct communication \cite{Thapliyal2018, jha2024joint}. This investigation forms the core of our study. {Some other experimental works show the use of multiphoton approach for several quantum communication setups.} Such as, \cite{teng2021sending} {proposes a modified, more efficient twin-field (TF) QKD using a sending-or-not-sending (SNS) TF-QKD, and also studies the case of two-photon emissions instead of single photon QKD.} Furthermore, \cite{biswas2022quantum} {studies coincidence measurements to monitor and selectively include multi-photon pulses (two- and three-photon events) in the key generation—rather than discarding them as in standard BB84—it is possible to boost the secure key rate by roughly 74\% over line-of-sight channels without resorting to decoy states.} \cite{mihaly2021effects} {studies the effect of different noise models on quantum memory of quantum repeaters which would be central to development of large scale quantum networks}. \cite{BisztrayBacsardi2018} {reviews the progress made in field of free-space QKD networks which highlights some of the key-experimental work done in deploying free-space QKD networks. Many of the fundamental works in field of QKD are theoretical and simulation based, while some provides experimental validation to these theoretical works. In this study, we simulate quantum networks based on several practical network parameters such as attenuation coefficient, noise parameters, etc. }

This paper looks at three different QKD protocols, i.e., the Decoy-state, the 3-stage, and the E91 protocols. Our study describes the efficiencies of the above protocols on different topologies such as direct, line, grid, ring, and torus topology. We vary network parameters such as entanglement swapping success probability, decoherence probability, and signal attenuation during transmission. We move on to analyze in detail the performance of E91 and the 3-stage protocol on the torus topology. To establish the significance of the multi-photon bursts in current practical scenarios, we analyze multi-photon bursts up to a burst size of a million qubits. The multiphoton burst represents generating a burst of a given size, and then encoding them as a quantum state. This allows for measurements, and thus collapsing the final result into on classical bit. This analysis led to defining a mathematical relationship between the size of the multi-photon burst used and the maximum distance of stable transmission between Alice and Bob. The preprint of this paper is located at \cite{jha2024multi}.

\section{QKD Protocols}
\label{Sec: QKD Protocols}
QKD protocols can be classified based on the detection techniques used to retrieve the key information encoded in the photons being used. Discrete-Variable (DV) protocols use the polarization (or phase) of weak coherent pulses to encode the information, which simulates a true single-photon state \cite{diamantinature}. Protocols such as Decoy-state and BB$84$ use the single photon-encoding scheme where the information is encoded in the polarization of the photon being used. Another category of protocols is called Continuous-Variable QKD, a technique where photon-counters are replaced with general p-i-n photo-diodes, which are known to be faster \cite{scarani2009}. The detection techniques used in the above are based on \textit{``homodyne detection"}.

\begin{figure}[h!]
    \centering
    \includegraphics[width=\linewidth]{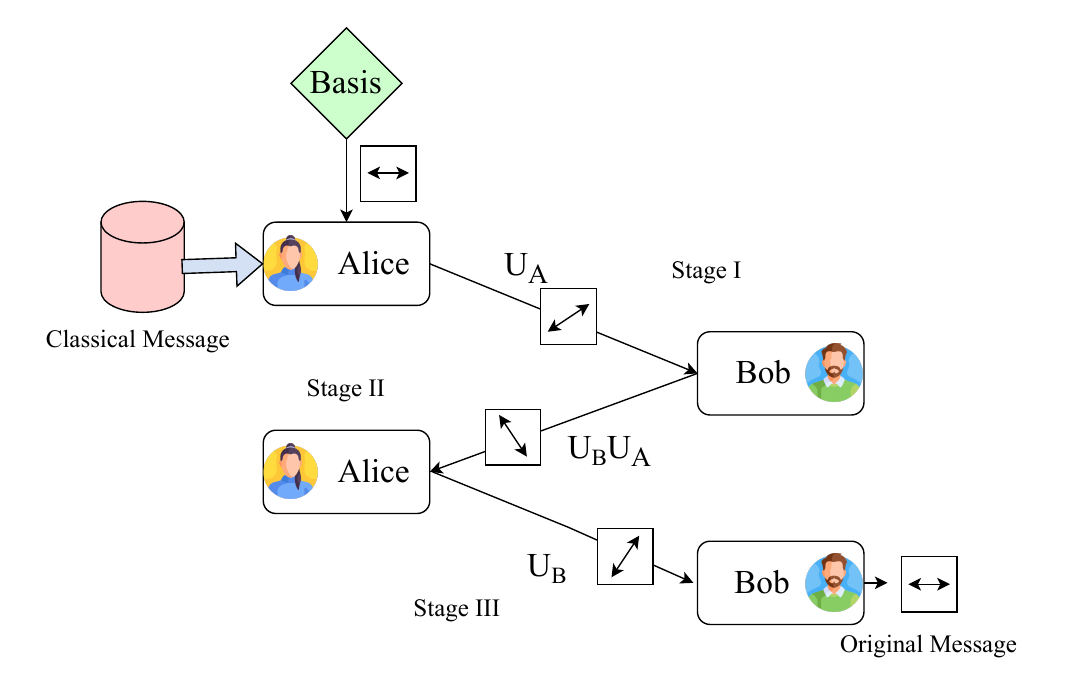}
    \caption{Schematic diagram representing the working of the three-stage protocol.}
    \label{fig:threestage}
\end{figure}

The \textit{Decoy-state} protocol enhances BB84 security by using random photon bursts to counteract PNS attacks, preventing Eve from distinguishing real from decoy transmissions \cite{bennett2014quantum}. Implemented using weak coherent pulses, this protocol varies the photon burst intensity and uses statistical analysis to detect intrusions \cite{lim2014concise, attema2021optimizing}. Post-transmission, Alice and Bob verify detection parameters over an authenticated channel to estimate and detect any attacker presence \cite{zhao2006simulation}. The \textit{three-stage} protocol is a viable option for practical quantum networks and it uses a technique similar to classical \textit{double-lock encryption}  \cite{Kak2006-3Stage, burr2022evaluating}. In this protocol, Alice starts by encoding a key or message using orthogonal quantum states or an arbitrary state $|X\rangle$. Both Alice and Bob apply secret unitary operations, $U_A$ and $U_B$, that commute (\([U_A, U_B] = 0\)) \cite{Kak2006-3Stage, burr2022evaluating}. The three-stage protocol is shown to be multiphoton resilient. This works by encoding multiple qubits for each classical bit, and since the preparation basis is global knowledge, all qubits associated with one bit are prepared in the same global basis. This multiphoton implementation of the three-stage protocol has been shown to be a viable option for sending direct messages across the quantum channel. \cite{mandal2013multi} Fig.(\ref{fig:threestage}) represents the working of the three-stage protocol introduced by \cite{Kak2006-3Stage}. The \textit{E91 protocol}, unlike the BB84 and B92 protocols, is an entanglement-based protocol utilizing single-photon transmissions. In our implementation, the $E91$ protocol uses the entanglement distribution through quantum repeater nodes through the entanglement distribution. {This can be interpreted as distributing multipartite entangled state across each pulse train. In post-processing, Alice and Bob keep only those coincidence events corresponding to one photon at each end; any higher‐order multiphoton coincidences are treated as erasures.}  This makes E91 more efficient over longer distances, as quantum repeaters help in the reduction of the attenuation loss. 

The main security concern with multi-photon transmissions is the PNS attack. We can define the probability of finding $n-$coherent photons as the Poisson distribution \cite{felix2001faint},
\begin{equation}
    P(n, \mu) = \frac{\mu^n e^{-\mu}}{n!},
    \label{poisson}
\end{equation}

where $\mu$ is the mean photon number for the photon burst. The zero, first, and second-order expansion for our photon distribution are,
\begin{equation}
    \begin{subequations}
        P(0) \approx 1-\mu +\frac{\mu^2}{2};\
    \end{subequations}
    \begin{subequations}
        P(1) \approx \mu-\mu^2;\
    \end{subequations}
    \begin{subequations}
        P(2) \approx \frac{\mu^2}{2}.\
    \end{subequations}
    \label{approx}
\end{equation}

Now, for a burst containing more than $1$ photon, i.e., $n \geq 2$, we can write the Poisson distribution for photons as \cite{felix2001faint},
\begin{equation}
    P(n\geq 2) \approx \frac{\mu}{2} + \frac{\mu^2}{4}.
    \label{second-order-approx}
\end{equation}

The two main strategies employed by the eavesdropper to compromise the security of quantum communications are \cite{felix2001faint}:
\begin{itemize}
    \item The first strategy involves Eve intercepting and analyzing all photons sent by Alice, and then relaying plausible states to Bob through a nearby source, carefully mimicking photon statistics. This scenario reveals that the maximum ratio of mutual information between Eve and Alice to the Quantum Bit Error Rate (QBER) reaches $6.83$. Under \textit{infinitesimal splitting}, it's shown that Eve could achieve complete informational equivalence with Alice, indicated by a mutual information score of $1$.
    \item The second strategy involves Eve using a beamsplitter to extract a fraction $\lambda$ from each pulse, simultaneously reducing photon loss by replacing the line with one of lower loss. The mutual information between Alice and Eve in this scenario is quantified as,
    \begin{equation}
        I(A,E) = \frac{\mu}{2}(2\lambda(1-\lambda))\frac{1}{2},
        \label{mutual-info}
    \end{equation}
    where $\mu/2$ represents the probability $P(2)$. This model predicts a maximum information gain for Eve of $\mu/8$ when $\lambda = 1/2$, correlating to a $3$ dB gain.
\end{itemize}


The key rate and the distance over which these key rates are stable are highly correlated to the presence of an eavesdropper in the system. The efficiency of the key distribution is highly dependent on the Quantum Bit Error Rate (QBER) and the mutual information between Alice and Eve, $I(A,E)$. With today's equipment, the major problem is the higher detector noise, which leads to high QBER at large distances; thus, the maximum distance of stable transmission drastically decreases \cite{felix2001faint}.

\begin{figure*}[ht!]
    \centering
    \begin{subfigure}[b]{0.3\textwidth}
        \includegraphics[width=\textwidth]{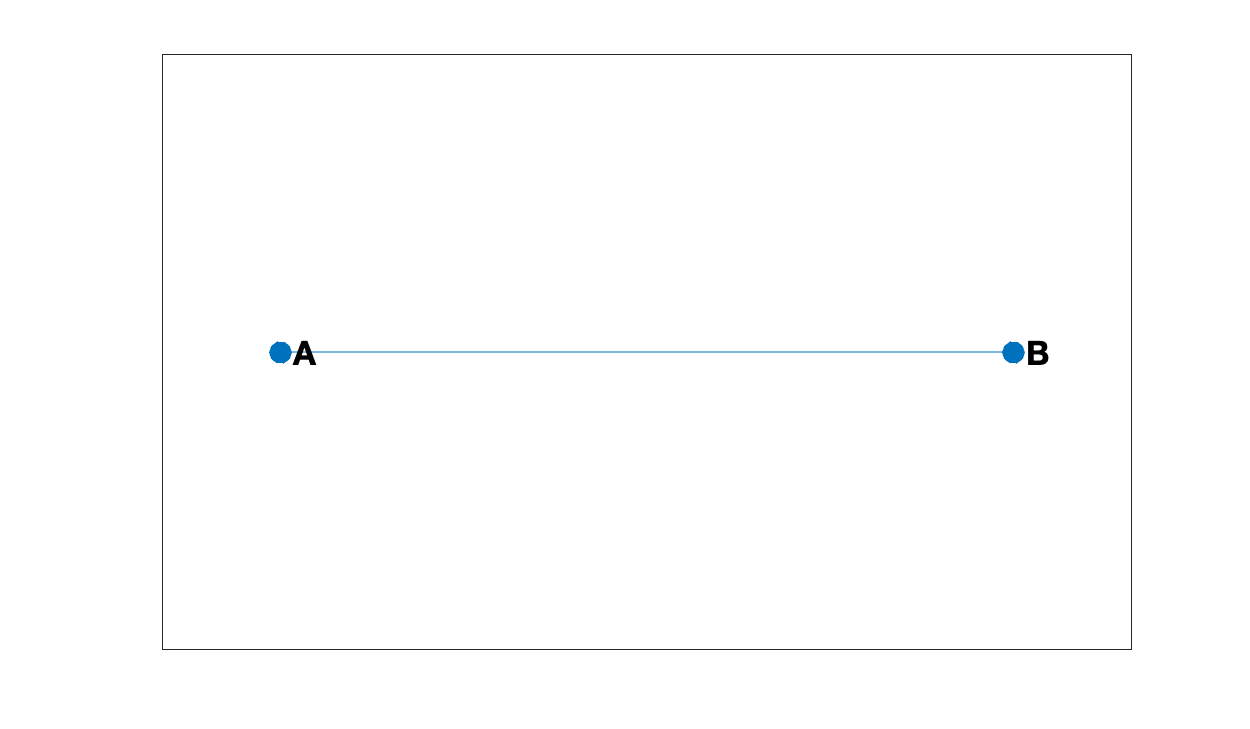}
        \caption{Direct Topology}
        \label{fig:Direct Topology}
    \end{subfigure}
    \hfill
    \begin{subfigure}[b]{0.3\textwidth}
        \includegraphics[width=\textwidth]{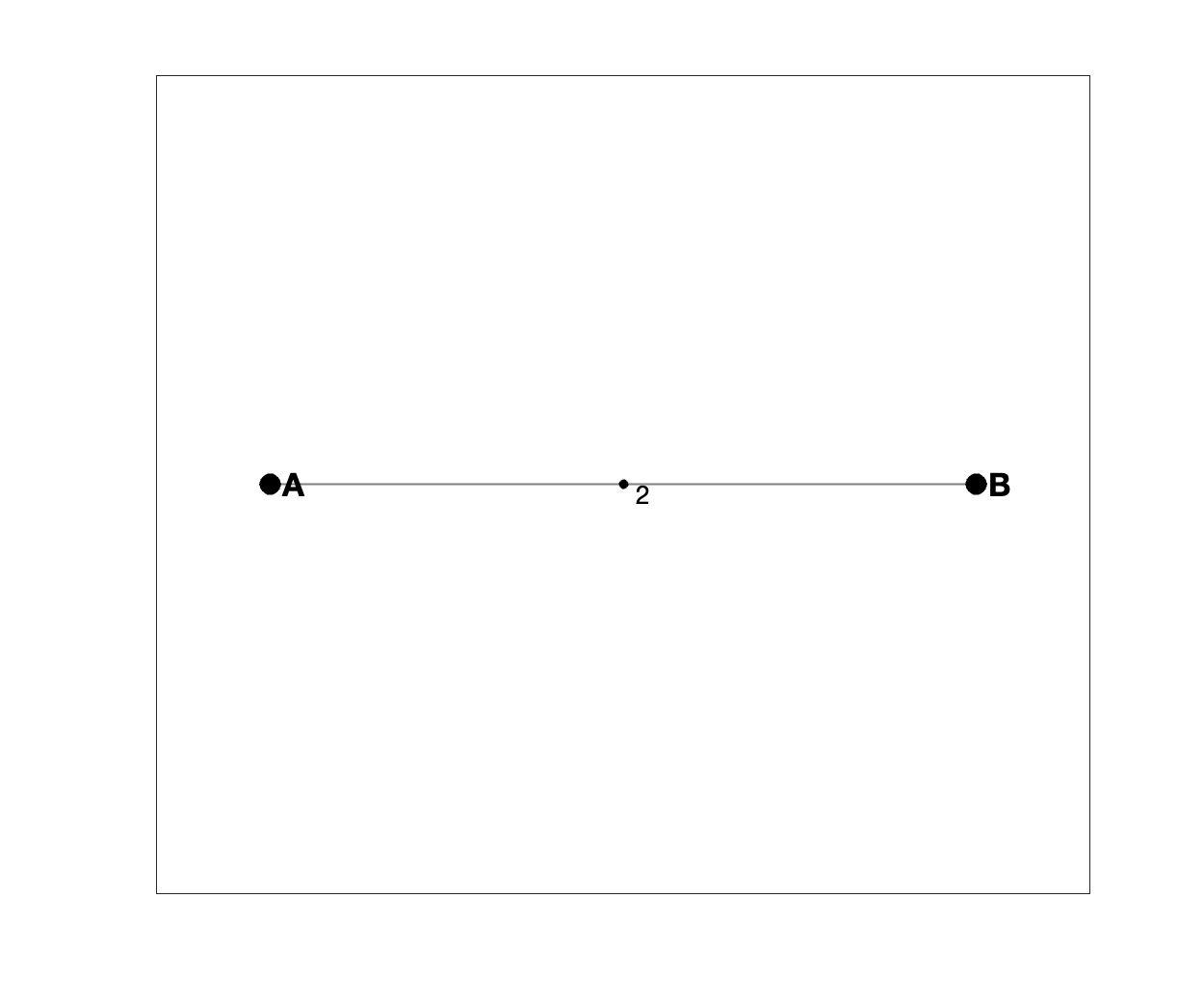}
        \caption{Line Topology}
        \label{fig:Line Topology}
    \end{subfigure}
    \hfill
     \begin{subfigure}[b]{0.3\textwidth}
        \includegraphics[width=\textwidth]{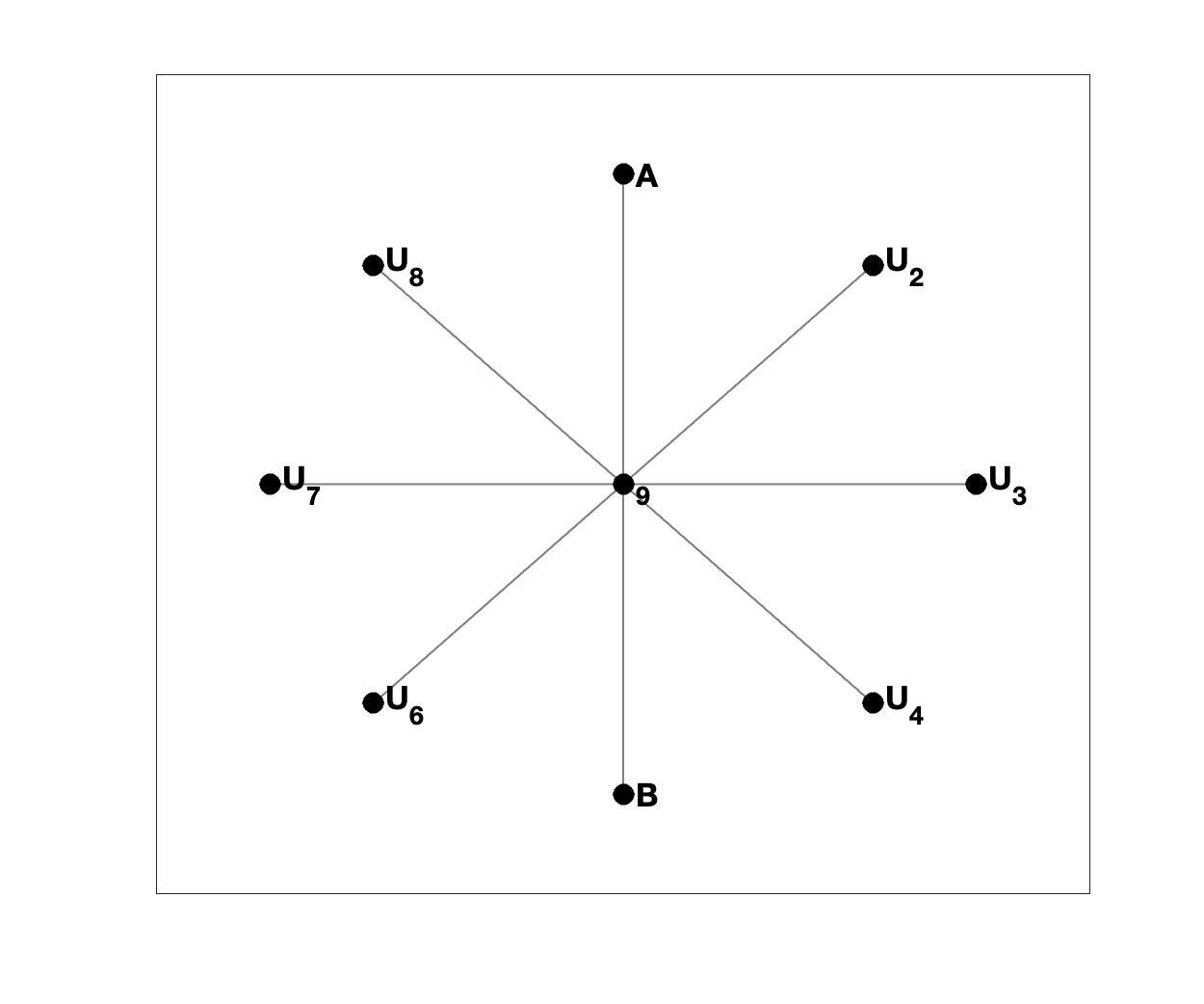}
        \caption{Star Topology}
        \label{fig:Star Topolgy}
    \end{subfigure}
    \hfill
     \begin{subfigure}[b]{0.3\textwidth}
        \includegraphics[width=\textwidth]{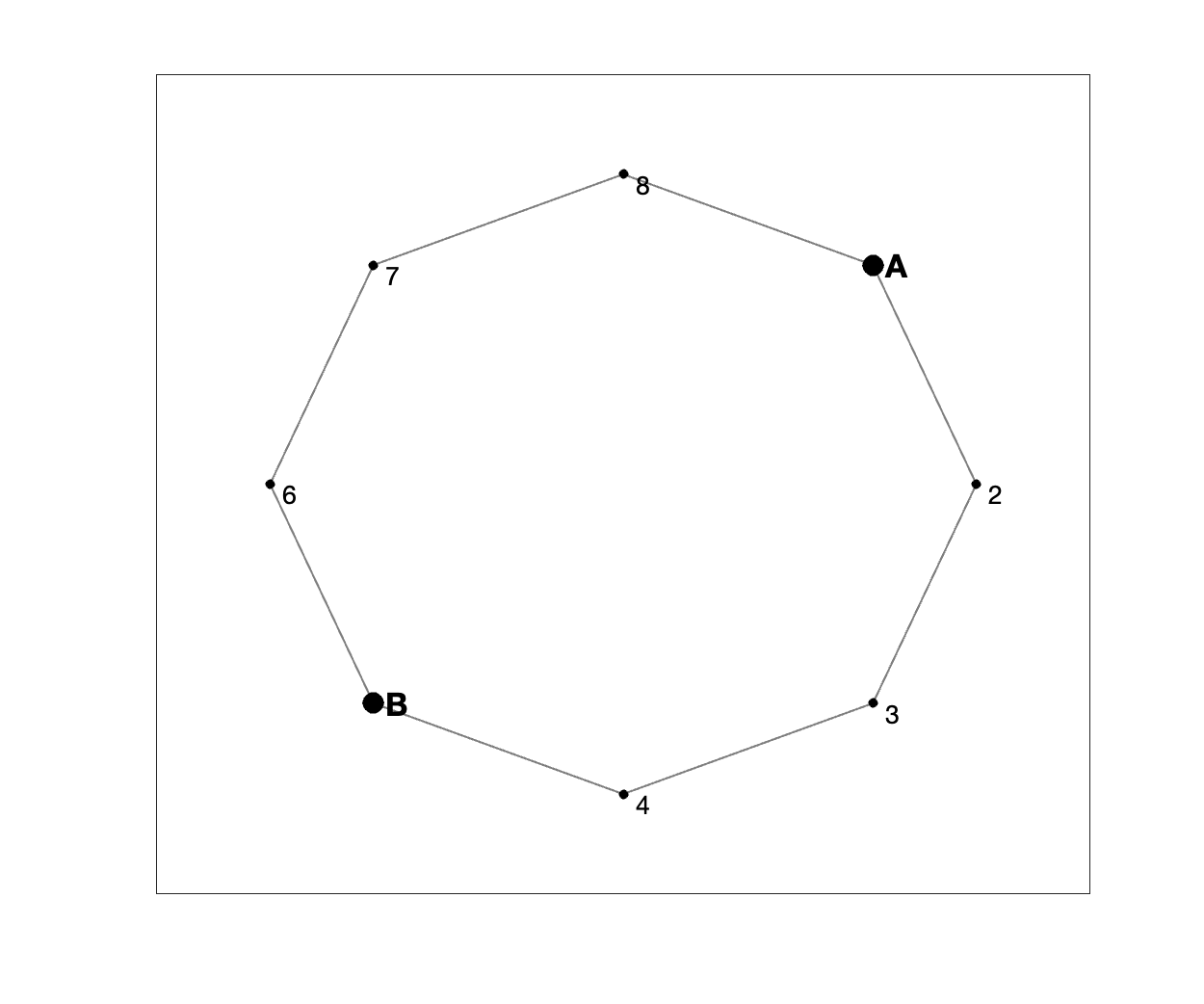}
        \caption{Ring Topology}
        \label{fig:Ring Topology}
    \end{subfigure}
        \hfill
     \begin{subfigure}[b]{0.3\textwidth}
        \includegraphics[width=\textwidth]{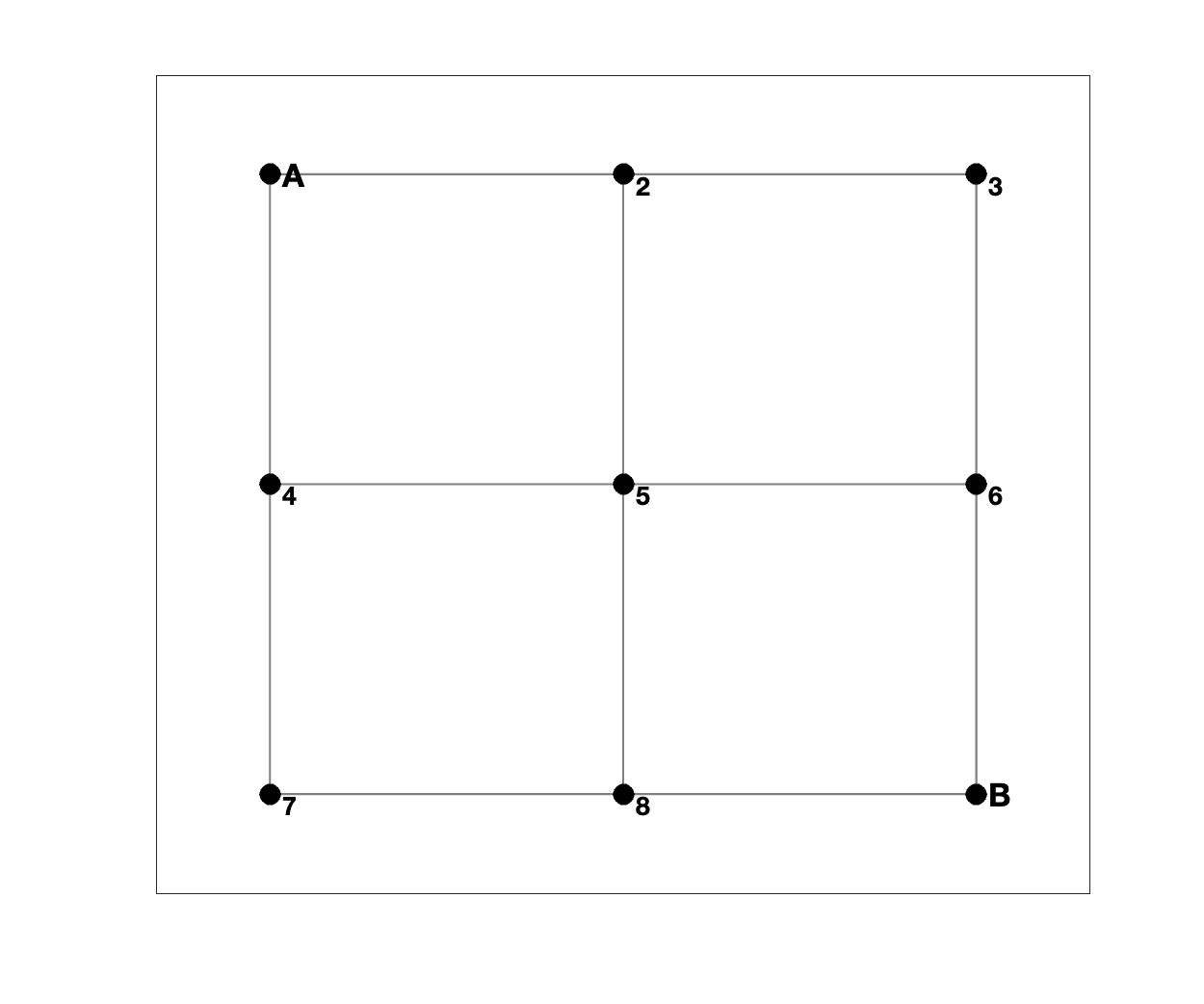}
        \caption{Grid Topology}
        \label{fig:Grid Topolgy}
    \end{subfigure}
        \hfill
     \begin{subfigure}[b]{0.33\textwidth}
        \includegraphics[width=\textwidth]{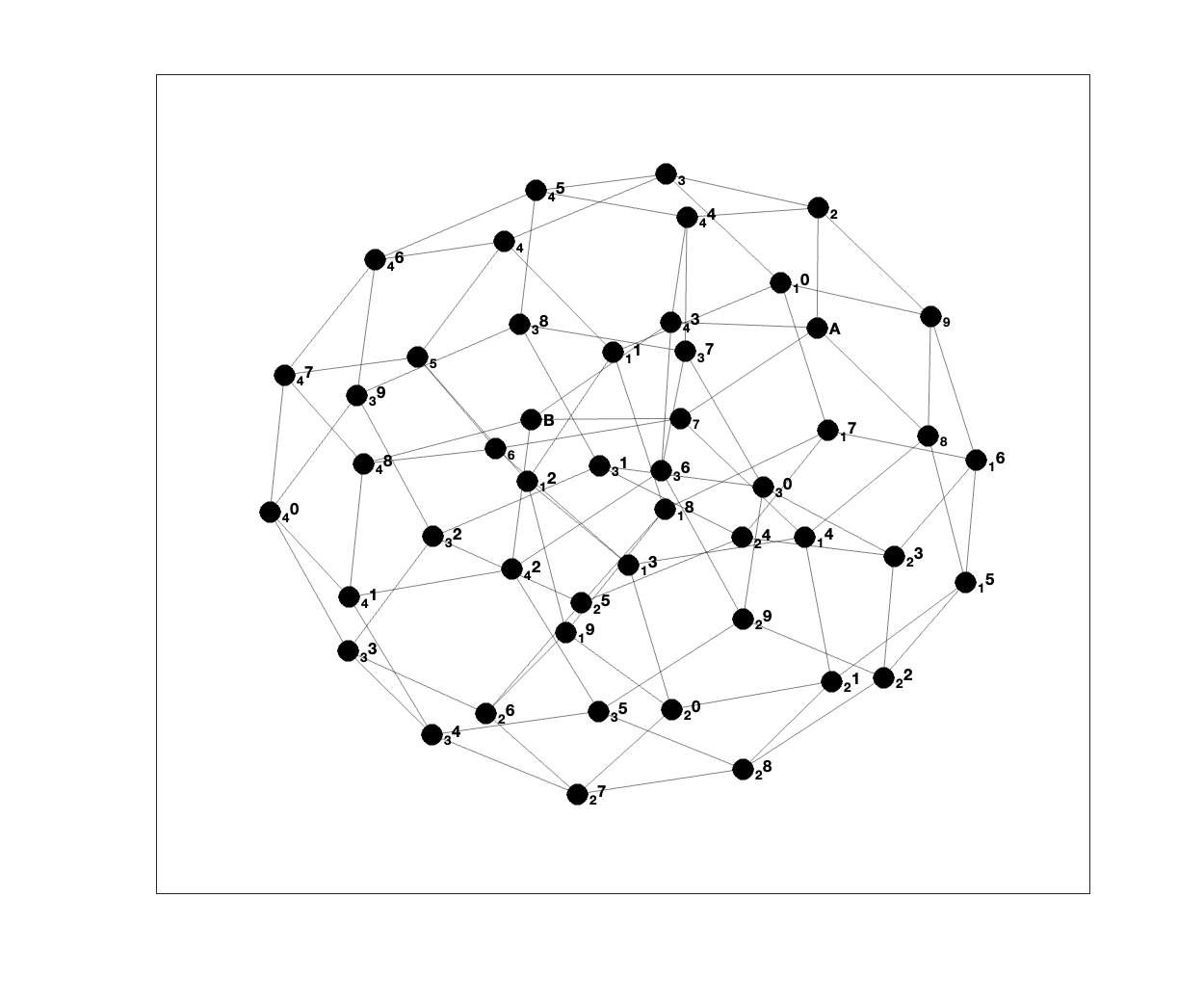}
        \caption{\(7\times 7\) Torus Topology. }
        \label{fig:Torus Topology}
    \end{subfigure}
    \vspace{0.2cm}
    \caption{Different topologies used in our QKD simulations (all of the topology profiles are generated through our simulations directly).}
    \label{QKD-Topologies}
\end{figure*}

One important aspect of developing efficient quantum networks is the efficiency of transmissions in different network topologies. Apart from studying some of the basic topologies, such as \textit{line}, \textit{star}, \textit{ring}, and \textit{grid}, we designed a \textit{torus topology} for our network simulator. Table \ref{tab:Topologies} compares the advantages and disadvantages of the various topologies used in this study. Figure (\ref{QKD-Topologies}) shows the different topologies generated through our network simulator.

\begin{table*}[ht]
\centering
\caption{Comparison of Different Topologies}
\vspace{0.2cm}
\label{tab:Topologies}
\begin{tabular}{|l|p{0.4\linewidth}|p{0.4\linewidth}|} 
\hline
Topology & Advantages & Disadvantages\\ \hline
Line & \vspace{0.4cm}Offers one path between Alice and Bob, i.e., minimum control-layered overhead & \begin{minipage}[t]{0.6\linewidth}
\vspace{0.03cm}
\begin{itemize}
    \item Impractical over long ranges.
    \item Not reliable.
\end{itemize}
\vspace{0.3cm} 
\end{minipage}\\ \hline
Gird & \begin{minipage}[t]{0.6\linewidth}
\vspace{0.1cm}
\begin{itemize}
    \item Provides multiple path between Alice and Bob.
    \item Nodes are geographically isolated.
\end{itemize}
\vspace{0.3cm}
\end{minipage} & \vspace{0.8cm}Maximum control-layer overhead needed. \\ \hline
Ring & Provides two paths between Alice and Bob with lesser control-layer overhead. & \vspace{0.03cm}Less reliable than grid topology. \\
\hline
Star & \vspace{0.03cm}Allows several user nodes at once. & Reduces overall performances and increased risk. \\ \hline 
Torus & Higher connectivity due to multiple available paths, and comparatively easier expansion without much reconstruction  & Higher initial and overall maintenance due to complex structure architecture.  \\ \hline
\end{tabular}
\end{table*}

\section{Network Simulation Setup}
\label{Sec:Simulator}
The Network Simulator developed for this study was written in Matlab. The network simulator was enhanced using the simulator presented in \cite{amer2020efficient}.

\subsection{Network Communication}
\begin{figure}[h!]
    \centering
    \includegraphics[width=0.4\textwidth]{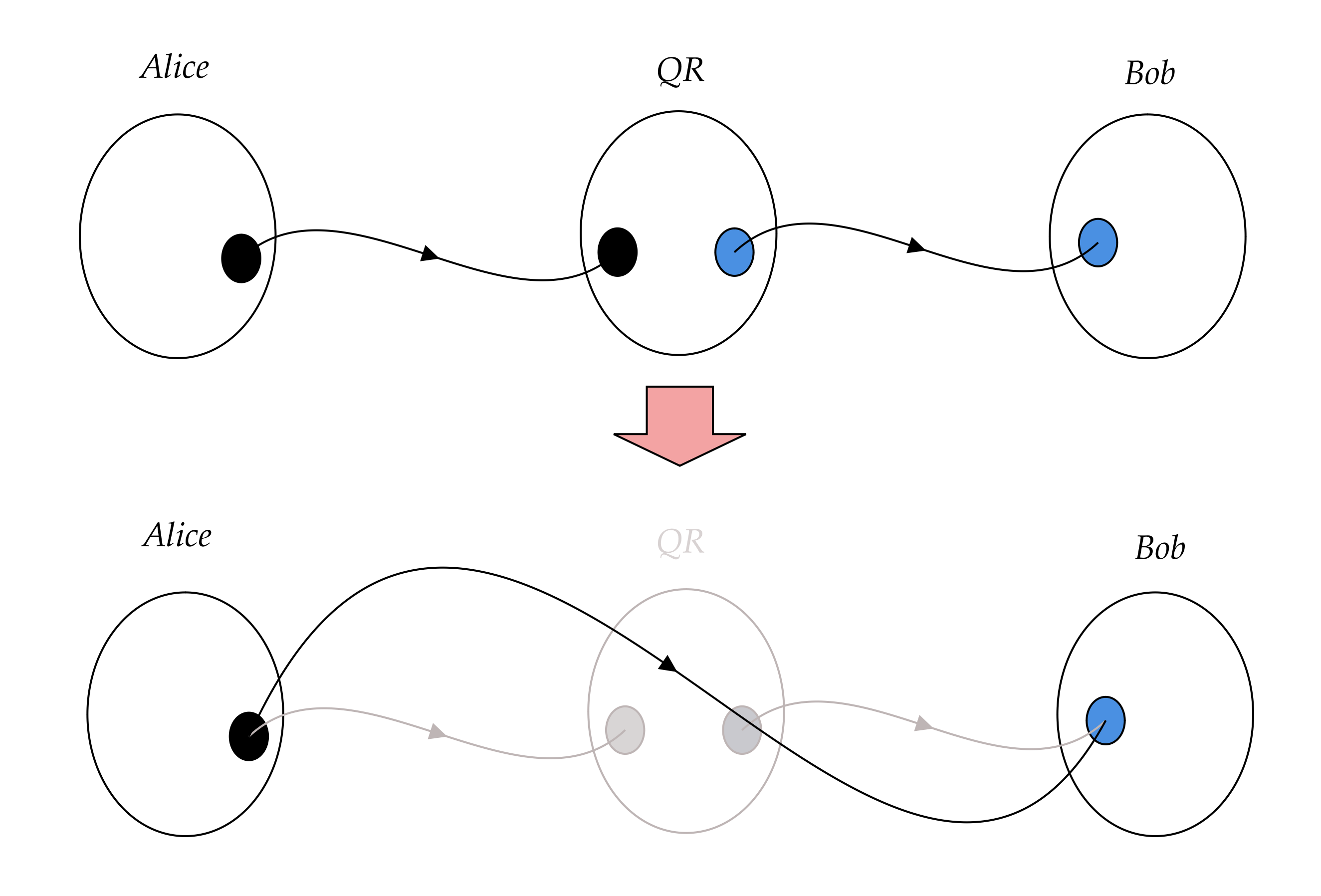}
    \caption{Working of the quantum repeater: Starts with two sets of entangled bell-pair between Alice-QR and QR-Bob. Then, we perform an entanglement swap between the nodes to create a final entangled Bell state pair between Alice and Bob. The curly line represents an entangled pair.}
    \label{entgswap}
\end{figure}

The connections between each of the involved nodes are achieved by simulating fiber-optic cables with transmission probability as given in equation(\ref{transprob}).
 
\begin{equation}
    P(L) = 10^{-\alpha L/10},
    \label{transprob}
\end{equation}

where $L$ is the fiber length in kilometers, and $\alpha$ is the attenuation coefficient of the fiber-optic cable used. It's evident from equation(\ref{transprob}) that as we increase the distance between the nodes or the length of fiber required, the success of transmission would drastically decrease. For our simulation, we keep the value of $\alpha = 0.15\ km^{-1}$ unless otherwise specified. Our network simulation includes various types of nodes, with trusted nodes like Alice and Bob being crucial. These nodes attempt to connect through viable paths, storing successfully transmitted qubits in a raw key pool, $(RK)_{i,j}$, where ${i,j}$ represents the node pair. The simulation also incorporates optical switches for linking nodes on a peer-to-peer basis, which, while facilitating dynamic connections, show performance deterioration. Further, we utilize the E91 Protocol to simulate quantum repeaters. Unlike classical repeaters, quantum repeaters use entanglement and entanglement swapping to extend the communication range without violating the no-cloning principle of quantum mechanics. The effectiveness of this method is shown in equation(\ref{transprob}), illustrating that although the probability of successful transmission between each repeater is high, the overall probability from Alice to Bob remains consistent as these events are independent. To address this, we simulate redundancy by conducting five simultaneous Bell state transmissions, enhancing entanglement swapping efficiency \cite{munro2010quantum}. Fig.(\ref{entgswap}) depicts the use of quantum repeaters in entanglement swapping using an external Bell pair through intermediate repeaters.

Once entanglement swapping is performed and the trusted nodes are connected, they proceed with the QKD protocol as if the qubits have been successfully transmitted.

\subsection{QKD Protocol Simulation}
In our QKD simulation, transmissions are completed using either quantum repeaters or direct fiber-optic cables as outlined in section \ref{Sec: QKD Protocols}. After conducting \(10^5-10^6\) QKD rounds, we simulate error correction factoring in a decoherence value of \(D=0.02\) from environmental noise impacting all qubits. We identify erroneous qubits, calculate the error rate \(Q\), and perform error corrections to derive the final key pool \(K_{i,j}\) between trusted nodes using equation(\ref{keypool}):
\begin{equation}
    K_{i,j} = R(1-2h(Q_{i,j})),
    \label{keypool}
\end{equation}
where \(h(Q)\) is the binary entropy function and $R$ is the raw-keyrate for the protocol run. Trusted nodes exchange key material through XOR operations. For example, node \(T_2\) assists in securely transmitting keys between Alice and Bob as shown in equation(\ref{xor}:
\begin{equation}
    K_{A,2} =  (K_{A,2}\oplus K_{2,B})\oplus K_{2,B}
    \label{xor}
\end{equation}
The overall key rate, which has the standard units of key-bits/s, indicating the QKD protocol's efficiency, is calculated by dividing the final key pool size by the number of QKD rounds.

\begin{table}[htbp]
\centering
\caption{Summary of quantum communication protocol simulation parameters.}
\vspace{0.2cm}
\label{tab:QKD Parameters}
\begin{tabular}{|l|p{0.7\linewidth}|} 
\hline
\textbf{Parameter} & \textbf{Description} \\
\hline
\( \alpha \) & Fiber attenuation coefficient. \( \alpha = 0.15 \). \( \alpha = 0.4 \) for optical switches.\\
\hline
\( L \) & Length of fiber segments in kilometers. \\
\hline
Burst Size & Frequency of photons in one burst. The default values are as follows,
\begin{itemize}
\item  Decoy State: A probability distribution between $0$ and $2$.
\item Three-Stage: $10$ photons sent in the first burst, then subsequent numbers.
\item E91: $5$ parallel Bell-state attempting between each quantum repeater.
\end{itemize}
\\
\hline
\( B \) & Probability of successful Bell state measurements for each of the quantum repeaters in between Alice and Bob. \( B = 0.85 \). \\
\hline
\( D \) & Probability of quantum state decoherence due to channel noise. \( D = 0.02 \). \\
\hline
\end{tabular}
\end{table}

\subsection{Torus Topology}
\label{Sec:Torus}
\begin{figure}[h!]
    \centering
    \includegraphics[width=0.75\linewidth]{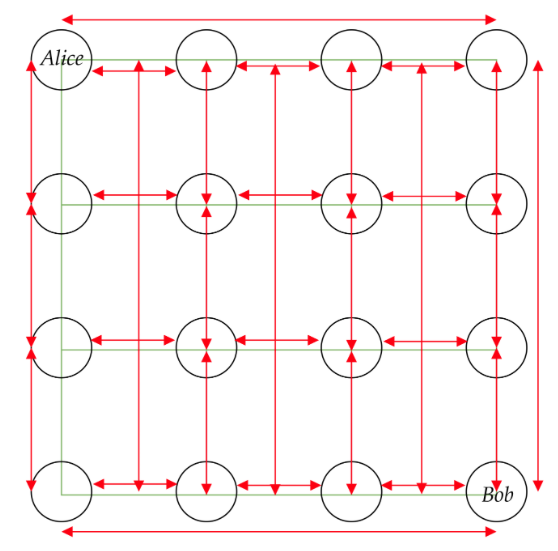}
    \caption{Schematic representation of the Torus topology. Red lines highlights the connection logic between user nodes. }
    \label{fig:torus-conn}
\end{figure}

In this study, we present a $3D$ topology for our network simulation along with several traditionally used topologies. Fig.(\ref{fig:Torus Topology}) shows the torus topology generated through the network simulator whereas Fig.(\ref{fig:torus-conn}) shows the schematic representation of the $2D$ representation of torus topology, highlighting the connection between the user nodes. The idea behind torus topology is a $3D$ wrapping of the connections in a traditional grid topology. 

\section{Results}
\label{Sec:Results}

In this section, we'll go over the results obtained through our simulations for different QKD Protocols as mentioned in section \ref{Sec: QKD Protocols} and over different Topologies.

\subsection{Performance over Direct Topology}
\label{Sec: Performance over Direct Topology}
\begin{figure}[h!]
    \centering
    \includegraphics[width=0.45\textwidth]{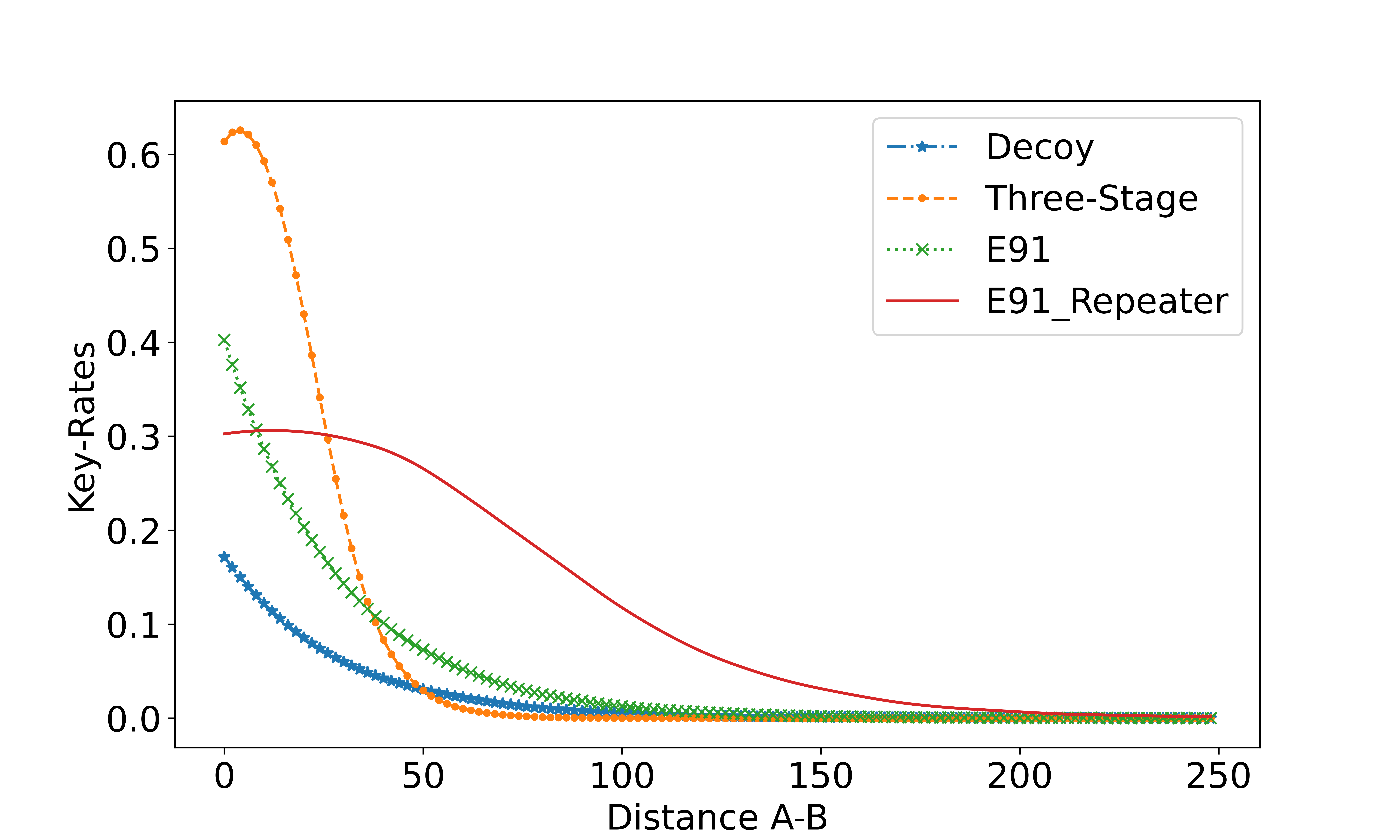}
    \caption{Comparison of Performance of different QKD Protocols over Direct Topology.}
    \label{fig:Direct QKD}
\end{figure}

We first looked at the performance of different QKD protocols in direct topology because it serves as the most basic case of topology without any complexities.  In this case, we simulated E91 in two cases, i.e., with and without using quantum repeaters. From Fig.(\ref{fig:Direct QKD}), we can see that using quantum repeaters significantly increased the distance of stable key rates to the case without any quantum repeaters. We see that three-stage protocol offers high key-rates for lower distances, however it reduces significantly with increasing distances. This can be associated with the fact that qubits have to travel thrice the distance than other protocols, thus attenuation loss is significantly higher. Decoy state offers the least key-rates due to vacuum bits occupying the bandwidth, and with similar attenuation, the number of effective qubits contributing to final key-rate is significantly lower. We see that E91 with repeater offers slightly lower key-rates, however it offers more stable key-rate over larger distances. 

\subsection{Performance of QKD Protocols over Different Topologies}
\label{Sec:Performance of QKD Protocols over Different Topologies}

\begin{figure*}[ht]
    \centering
    \begin{subfigure}{0.47\textwidth}
        \centering
        \includegraphics[width=\textwidth]{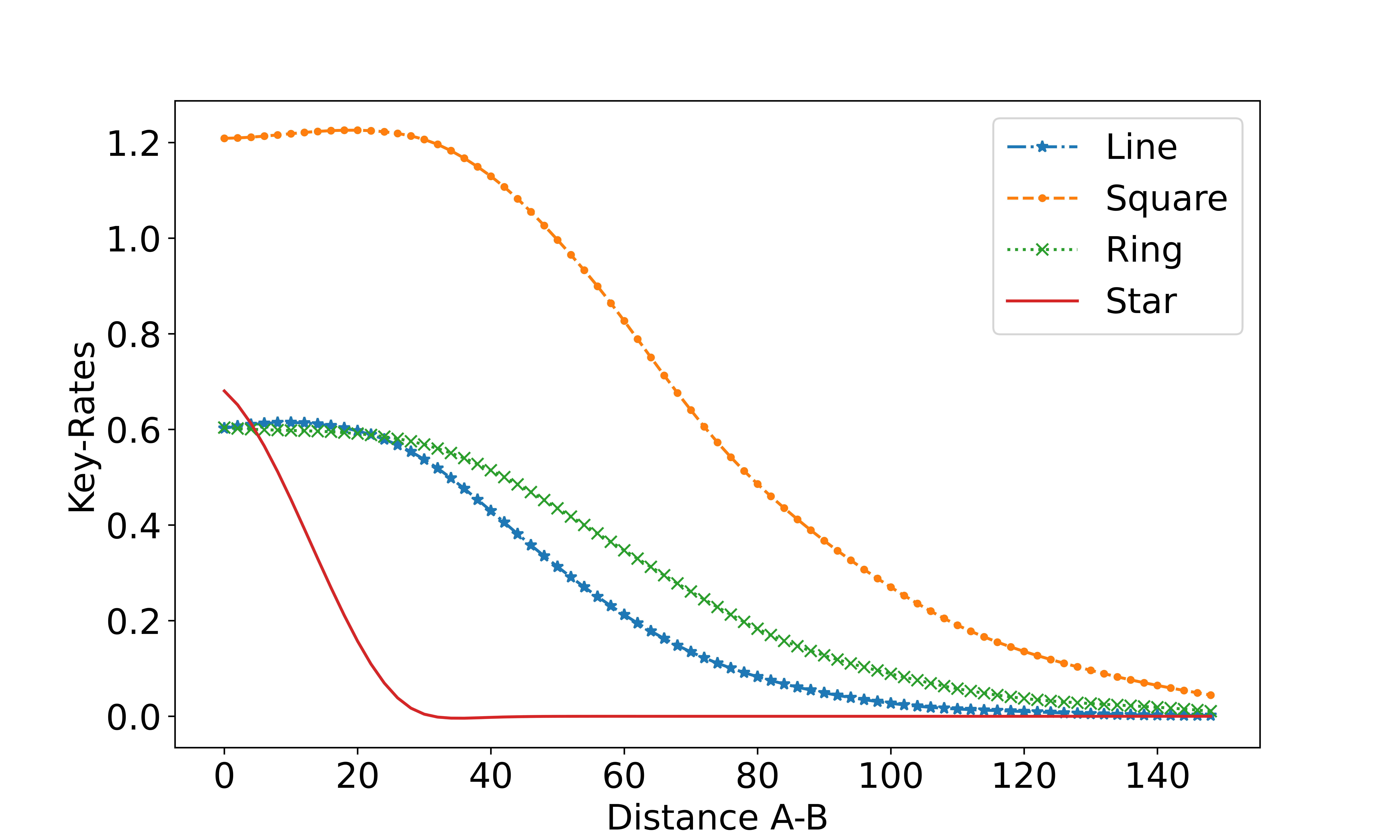}
        \caption{Three Stage Protocol Profile}
        \label{fig:Three Stage Protocol Profile}
    \end{subfigure}
    \begin{subfigure}{0.47\textwidth}
        \centering
        \includegraphics[width=\textwidth]{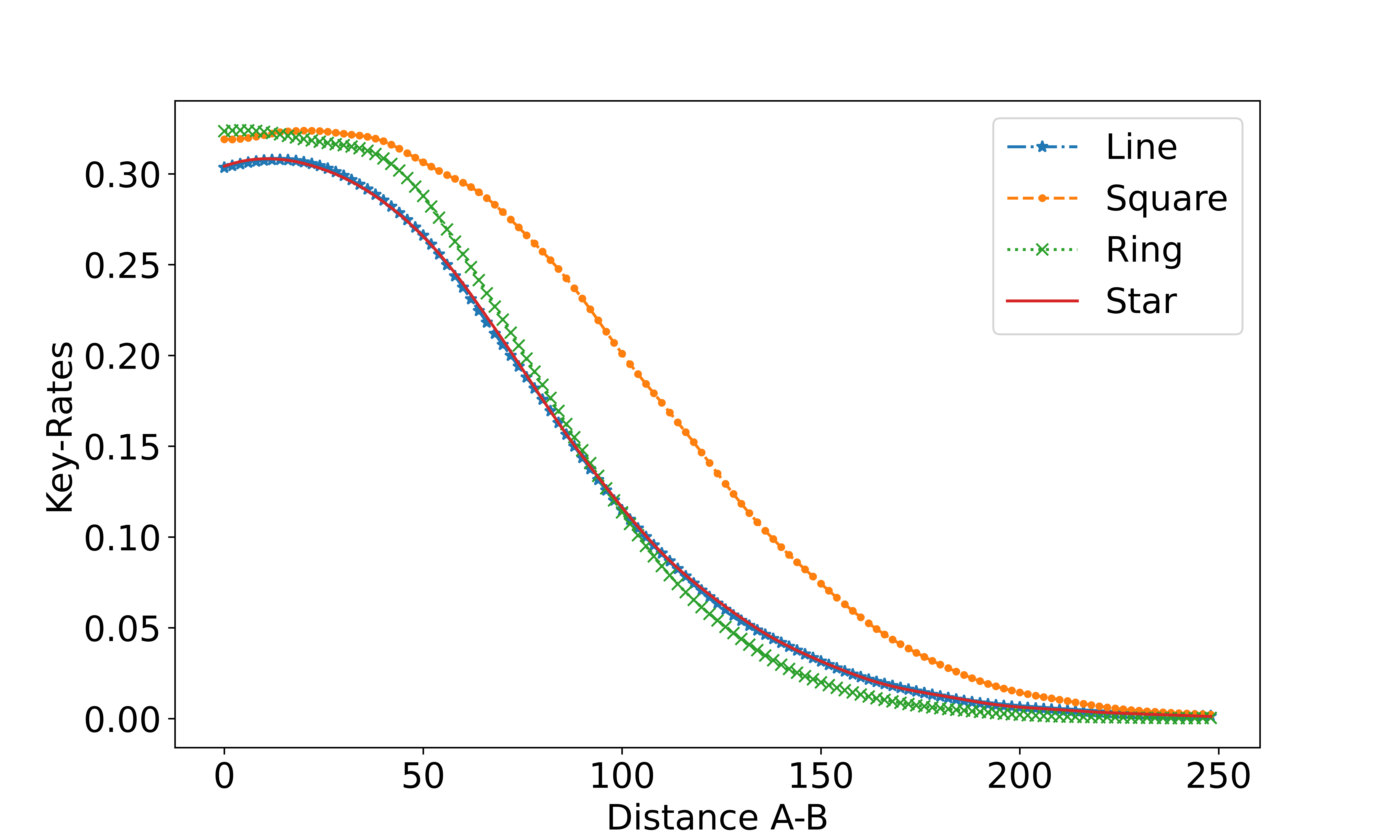}
        \caption{E91 Protocol Profile}
        \label{fig:E91 Protocol Profile}
    \end{subfigure}
    \begin{subfigure}{0.47\textwidth}
        \centering
        \includegraphics[width=\textwidth]{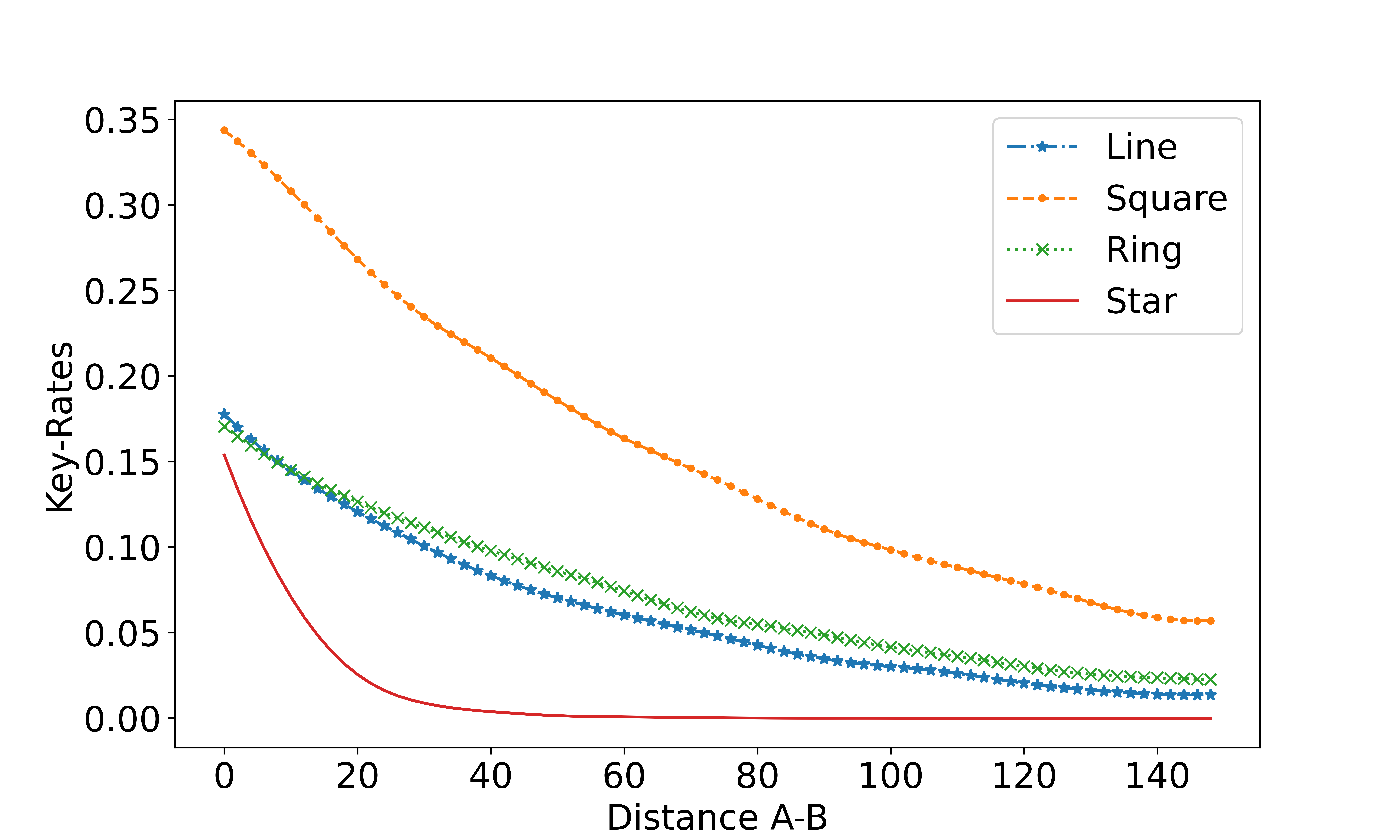}
        \caption{Decoy-State Protocol Profile}
        \label{fig:Decoy-State Protocol Profile}
    \end{subfigure}
    \caption{Performance of different QKD Protocols over Different Topologies}
    \label{fig:QKD-Topology}
\end{figure*}

This section explores the performance of different QKD protocols over different Topologies as described in Fig.(\ref{QKD-Topologies}). We compare the changing key rates over various distances between Alice and Bob for the 3-stage protocol, the E91 protocol, and decoy-state protocol (section \ref{Sec: QKD Protocols}). As seen in Fig.(\ref{fig:QKD-Topology}), Decoy-State does not offer very stable transmission distances, and the decay is more rapid than the other two protocols. One other interesting thing to note is that the grid topology (as shown in Fig.(\ref{fig:Grid Topolgy}) offers significantly higher key rates for 3-stage protocol due to the availability of multiple paths for key distribution, i.e., the possibility of more than one key being distributed each round because of the presence of multiple trusted nodes. 

\subsubsection*{QKD Protocol Performance over Torus Topology}
\label{Sec: QKD Protocol Performance over Torus Topology}
This section explores the performance of the 3-stage protocol and E91 protocol over our torus topology. As we can see from Fig.(\ref{fig:Torus Topology}), there are multiple paths between Alice and Bob, and unlike grid topology (Fig.(\ref{fig:Grid Topolgy})), the shortest path between Alice and Bob is $L$ itself. This increases the probability of successful transmission between Alice and Bob, and thus, we can expect a higher key rate than that of the grid topology. As shown earlier in Fig.(\ref{fig:Direct QKD}) the key rates were observed to be higher for the 3-stage Protocol than the E91 Protocol. We see similar behavior from Fig.(\ref{char-curves}) that the 3-stage Protocol beats the E91 Protocol almost by 1.5 times. We also see that the stable distance for E91 is observed to be higher, as seen in the previous cases as well.
\begin{figure*}[h!]
    \centering
    \begin{subfigure}[t]{0.47\textwidth}
        \includegraphics[width=\textwidth]{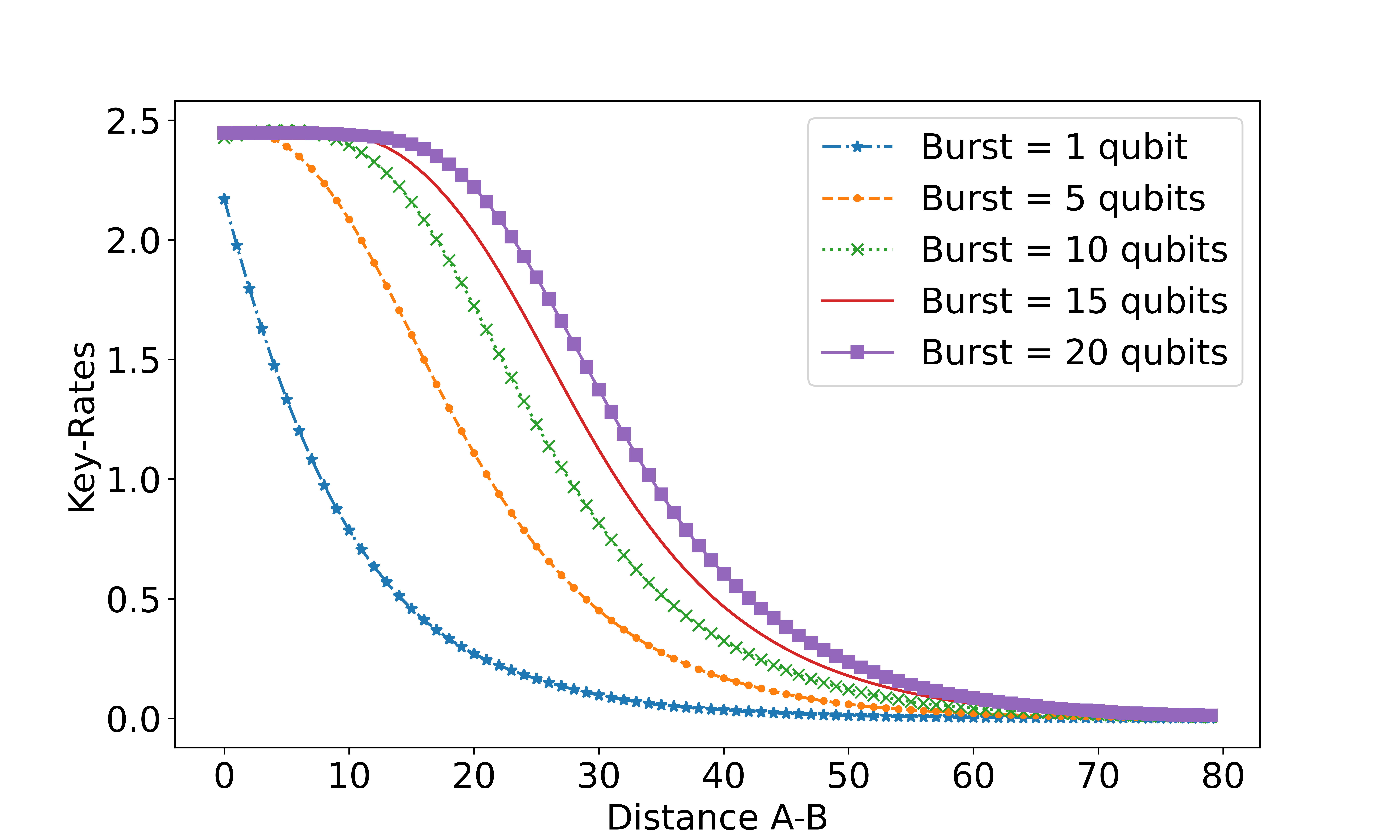}
        \caption{3- Stage QKD Protocol Performance over a $3\times 3$ Torus Topology.}
        \label{fig:torus3stage}
    \end{subfigure}
    \hfill
    \begin{subfigure}[t]{0.47\textwidth}
        \includegraphics[width=\textwidth]{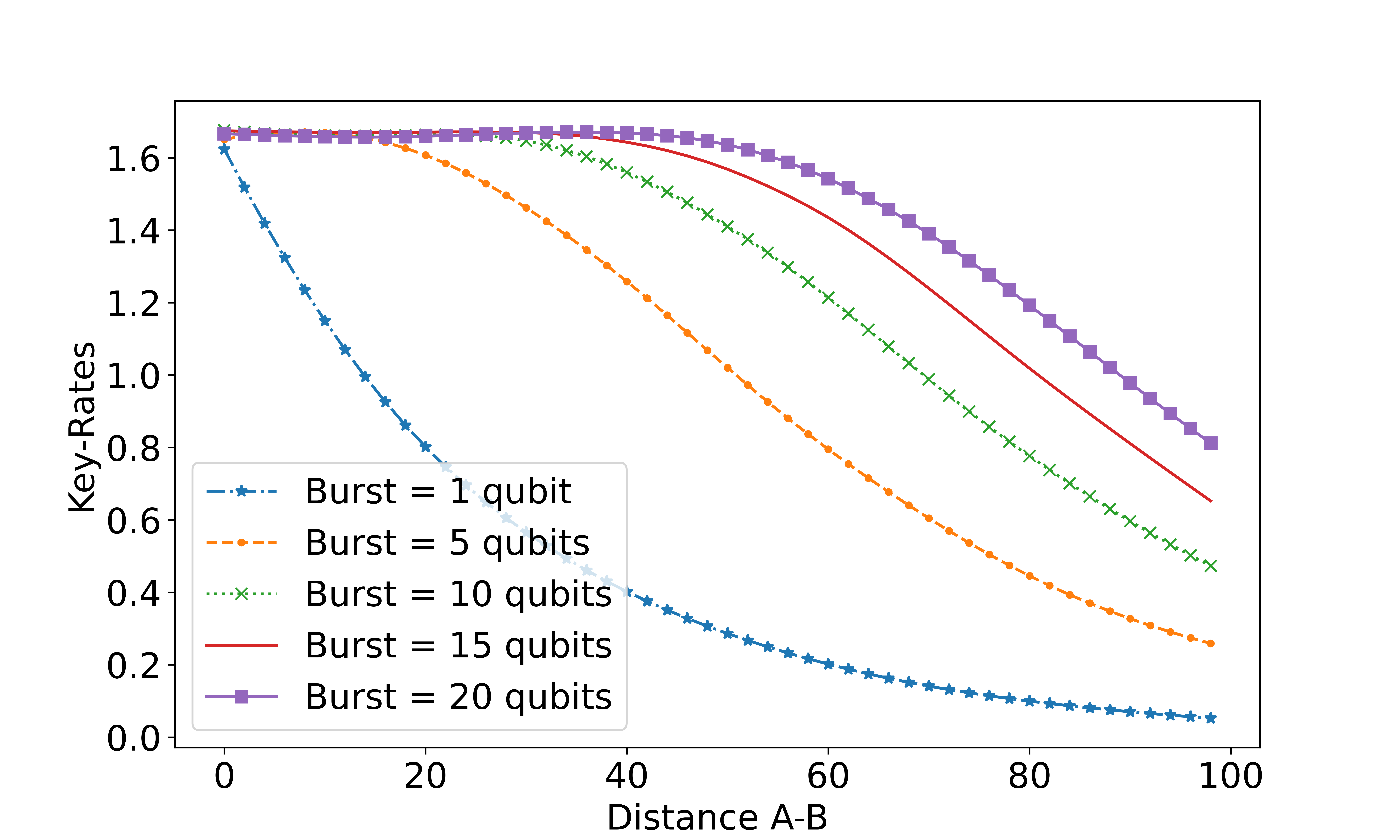}
        \caption{E91 QKD Protocol Performance over a $3\times 3$ Torus Topology.}
        \label{fig:toruse91}
    \end{subfigure}
    \hfill
    \vspace{0.2cm}
    \caption{3-Stage and E91 Protocol Performance over Torus Topology.}
    \label{char-curves}
\end{figure*}

\subsection{Performance of the Decoy-State Protocol}
\label{Sec: Performance of Decoy-State Protocol}
From Fig.(\ref{fig:Decoy-State Protocol Profile}), we can see that the grid topology offers the highest key rates than the rest of the topologies due to multiple paths and multiple trusted nodes contributing to more than a single key per simulation round. The grid topology also seems to be more durable over time because the fiber segments between trusted nodes are shorter compared to other topologies used. Ring topology also offers multiple paths, but it seems to be just slightly better than line topology. However, over time, ring topology seems to show a better key rate with a lower decline rate than the line topology. For star topology, the higher attenuation coefficient ($\alpha = 0.4$ because of optical switches) results in significantly more fiber loss than any other topologies as described by equation(\ref{transprob}).
\subsection{Performance of the Three-Stage Protocol}
\label{Sec: Performance of Three-Stage Protocol}
Figure \ref{fig:Three Stage Protocol Profile} indicates that the grid topology consistently achieves higher key rates due to its multiple paths and trusted nodes, enhancing robustness and key pool size. Contrasting the 3-stage protocol's performance with direct and line topologies, as shown in figures Fig.(\ref{fig:Direct QKD}) and Fig.(\ref{fig:Three Stage Protocol Profile}), the line topology exhibits slower key-rate decline over longer distances due to shorter fiber segments and a higher probability of successful transmission (equation(\ref{transprob}). At shorter distances, line and ring topologies perform similarly, but the ring topology proves more durable at greater ranges, showing less key-rate decay. Conversely, the star topology, despite starting with higher key rates, experiences rapid decay because of increased attenuation from simulating an optical switch.
\subsection{Performance of E91 Protocol}
\label{Sec: Performance of E91 Protocol}
Fig.(\ref{fig:E91 Protocol Profile}) demonstrates that while most topologies show similar performance under the E91 protocol, the grid topology shows robustness and slower key rate decay, attributed to its multiple paths and three repeaters in the shortest path. The ring topology, in contrast, achieves higher key rates at very short distances due to fewer repeaters, reducing the likelihood of entanglement swapping failures. Despite its benefits, the grid topology's advantage is offset by errors from failed quantum repeaters. Conversely, the star topology leverages a central repeater to directly connect any two nodes by creating Bell states selectively, circumventing the need for an optical switch and matching the efficiency of direct line topologies that rely on a single quantum repeater for connectivity between nodes like Alice and Bob.
\subsection{Analysing Higher Order Multi-Photon Bursts}
\label{Sec:Analysing Higher Order Multi-Photon Bursts }
In this section, we explore the trade-offs between performance and security risks with larger multi-photon bursts in the 3-stage protocol, which previously utilized a default burst size of 10 qubits. While increasing burst size can enhance performance, it also raises the potential for PNS attacks. For the E91 protocol, which does not naturally support multi-photon bursts, performance adjustments involve varying the number of simultaneous Bell pairs between quantum repeaters.

\begin{figure*}[h!]
    \centering
    \begin{subfigure}[t]{0.48\textwidth}
        \centering
        \includegraphics[width=\textwidth]{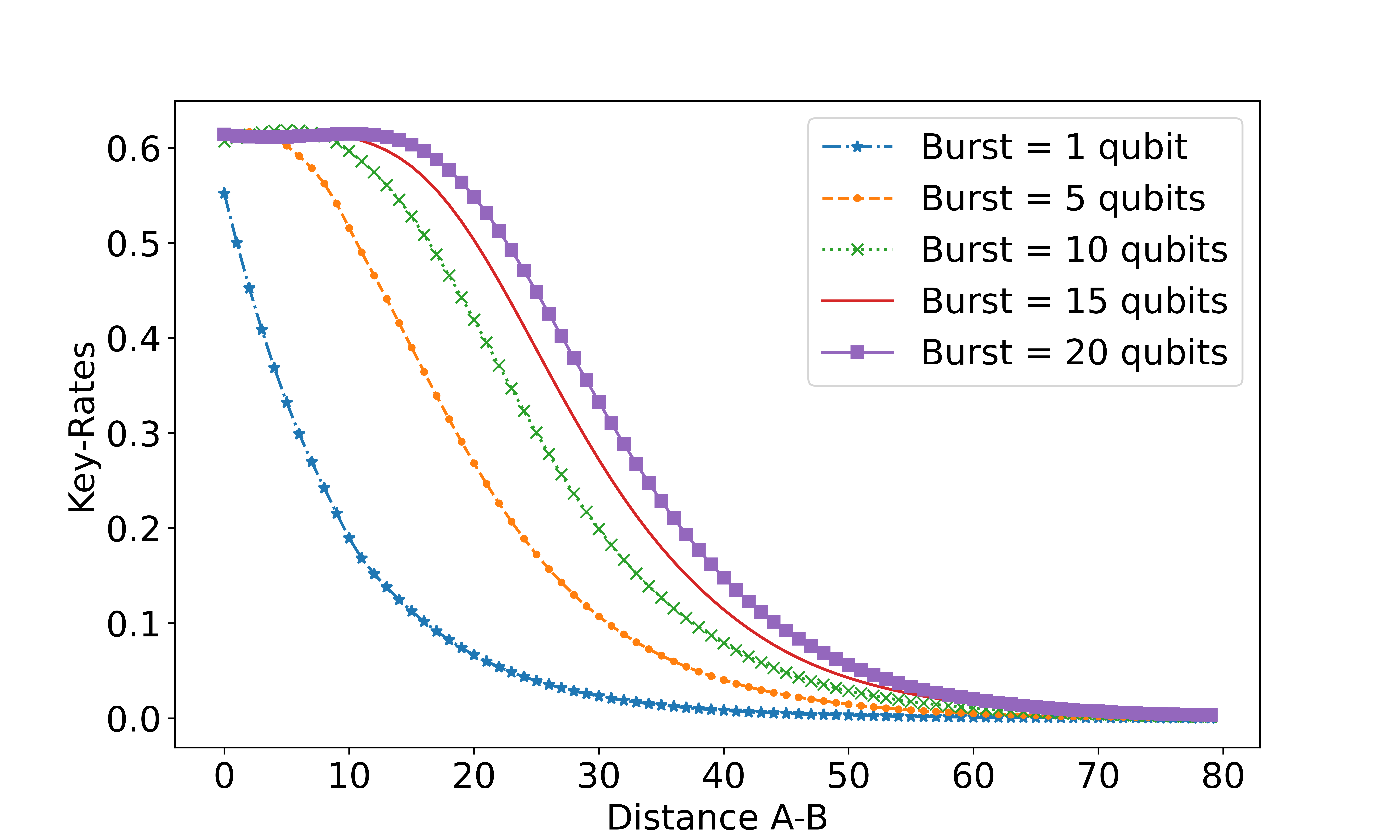}
     \caption{Multi-Photon Burst Profile for Three-Stage Protocol over Line Topology.}
        \label{multiphoton-3stage}
    \end{subfigure}
    \begin{subfigure}[t]{0.48\textwidth}
        \centering
        \includegraphics[width=\textwidth]{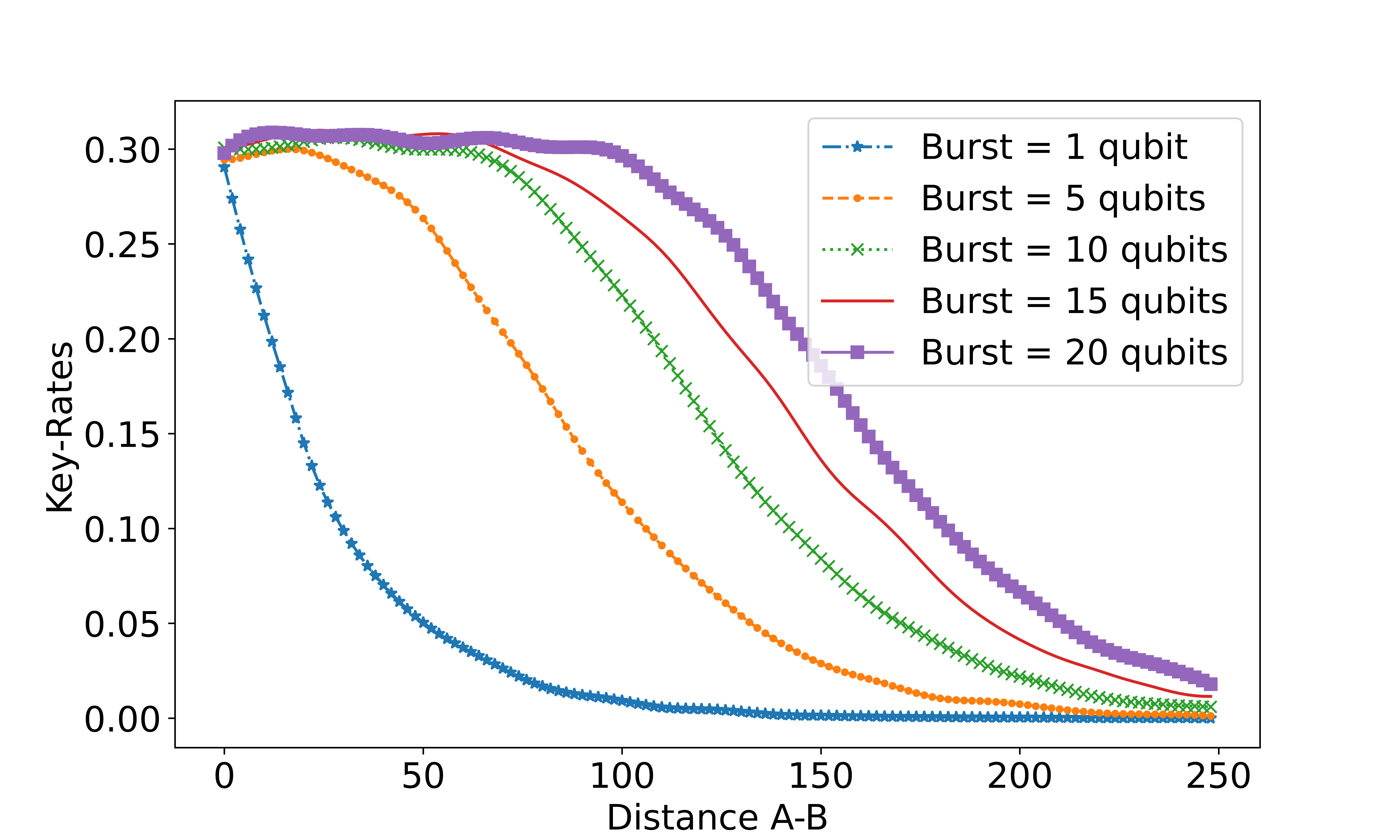}
            \caption{Multi-Photon Burst Profile for E91 over Line Topology.}
            \label{multiphoton-e91}
    \end{subfigure}
    \hfill
    \caption{Comparison of Multi-Photon Burst Size Profiles for Three-stage and E91 Protocol}
    \label{fig:Multiphoton-burst}
\end{figure*}
Fig.(\ref{multiphoton-3stage}) demonstrates that larger burst sizes generally maintain stable key rates at shorter distances for the 3-stage protocol. Meanwhile,  Fig.(\ref{multiphoton-e91}) shows that the E91 protocol sustains longer-range stable transmissions than the 3-stage protocol, as corroborated by Figures  Fig.(\ref{fig:Direct QKD}) and Fig.(\ref{fig:E91 Protocol Profile}). However, both protocols see key rates diminish at extended distances, with variations depending on the burst size and protocol. These findings highlight the importance of exploring various network topologies and integrating multiple trusted nodes to enhance protocol efficacy.
\subsection{Multi-Photon Burst and Distance Relation}
\label{Sec: Multi-Photon Burst and Distance Relation}
This section looks into finding a mathematical relation governing these curves to define a predictive model for the simulations. We notice a few features from the curves, 
\begin{itemize}
    \item A stable flat curve for smaller distances between Alice and Bob.
    \item A decaying curve, sort of linear, in the middle section. 
    \item The curve converges to zero at larger distances. 
\end{itemize}
We define an exponentially decaying function of the sigmoid nature as described in equation(\ref{sample-eq}),
\begin{equation}
    y = \frac{R}{(1+e^{k(x-x_0)})},
    \label{sample-eq}
\end{equation}
{where, $R$, $k$, $x_0$ are the fitted parameters, and $x$ is the inter-node distance between Alice and Bob (i.e., the sender and the receiver)}. $R$ defines the initial constant value of key-rate, $k$ defines the decaying rate of the curve, and $x_0$ is the point of the curve where key-rate $=R/2$. We do curve fitting for the line and ring topologies and present the results below in Fig.(\ref{fig:curve-fit}).

\begin{figure*}[ht!]
    \centering
    \begin{subfigure}[t]{0.4\textwidth}
        \includegraphics[width=\textwidth]{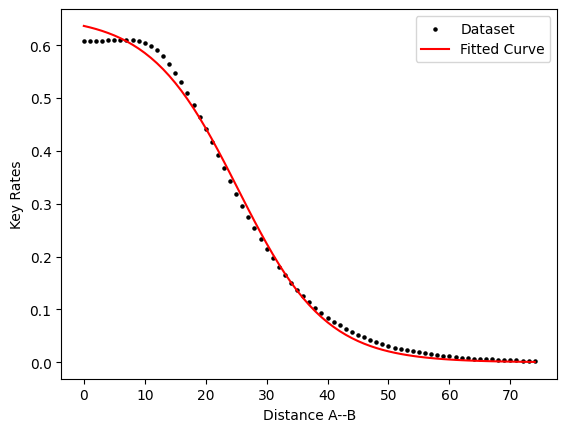}
        \caption{Characteristic curve obtained for QKD Performance over a Line-topology.}
        \label{fig:LINE-TOPO-MP}
    \end{subfigure}
    \hfill
    \begin{subfigure}[t]{0.4\textwidth}
        \includegraphics[width=\textwidth]{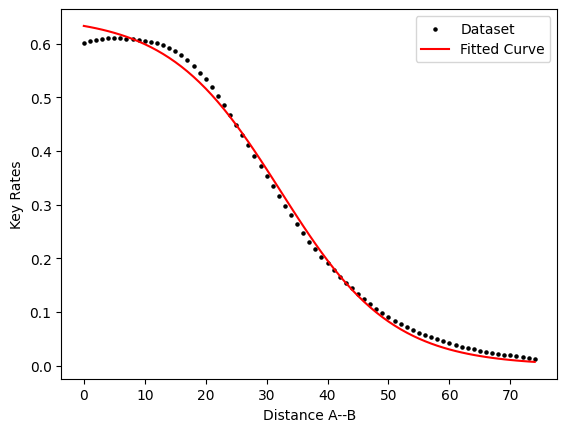}
        \caption{Characteristic curve obtained for QKD Performance over a Ring-topology.}
        \label{fig:RING-TOPO-MP}
    \end{subfigure}
    \vspace{0.2cm}
    \caption{Fitting the curve to the equation(\ref{sample-eq}) for Line and Ring topology for a multi-photon burst size of $10$ qubits for 3-stage QKD Protocol.}
    \label{fig:curve-fit}
\end{figure*}

Based on the values of various fitted parameters from equation(\ref{sample-eq}), the following were the equations of the curves found for the above are, 

\begin{equation}
    \begin{subequations}
        y_{\text{line}} = \frac{0.655}{(1+e^{0.139(x-25.303)})}\
    \end{subequations}
    \ \text{and}\ \
    \begin{subequations}
        y_{\text{ring}} = \frac{0.652}{(1+e^{0.109(x-32.257)})}
    \end{subequations}
    \label{eq:fit-curve}
\end{equation}

\noindent equation(\ref{eq:fit-curve}) describes the characteristic equations for line and ring topology for $3-$stage protocol.

\subsubsection{Multi-Photon Burst Profiles}
\label{Sec: Multi-Photon Burst Profiles}

This section explores the multi-photon burst relations for higher order bursts over line topology for the 3-stage protocol. We explore the key rates and distance relation for the following burst sizes, $b = [50, 1200]$. We first make the curve smooth by eliminating the polynomial noise using the SavGol filter. We then fit the curves using equation(\ref{sample-eq}) and present the characteristic curves for each of them in Fig.(\ref{char-curves}).

\begin{figure*}[h!]
    \centering
    \begin{subfigure}[t]{0.4\textwidth}
        \includegraphics[width=\textwidth]{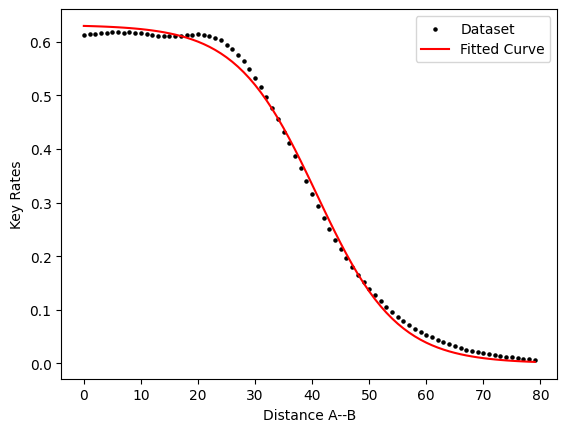}
        \caption{Characteristic curve obtained for burst size $=50$ qubits for 3-stage protocol.}
        \label{fig:50 qubits}
    \end{subfigure}
    \hfill
     \begin{subfigure}[t]{0.4\textwidth}
        \includegraphics[width=\textwidth]{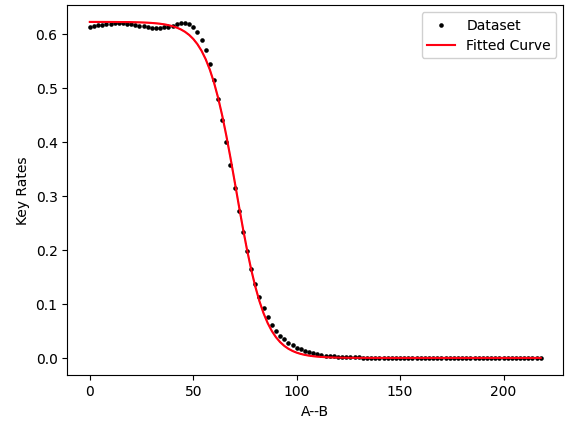}
        \caption{Characteristic curve obtained for burst size $=1200$ qubits for 3-stage protocol.}
        \label{fig:1200 qubits}
    \end{subfigure}
    \vspace{0.2cm}
    \caption{Fitting the curve to the equation(\ref{sample-eq}) for line topology having higher order multi-photon bursts used for 3-stage Protocol.}
    \label{char-curves}
\end{figure*}

For lower order burst size, we see the fit of the curve being better as key rates approaches zero (for Fig.(\ref{fig:50 qubits})), and for higher order curves (such as Fig.(\ref{fig:1200 qubits})) we can clearly see that the curve fit becomes better for the initial region as well. One important thing to note is that we can see the constant part of the curve increases with the increasing multi-photon burst size; therefore, we need to establish a relationship between the two quantities as well. This will be done in the next section. 

\subsection{Distance of Stable Transmission and Multi-Photon Burst Size}
\label{Sec: Distance of Stable Transmission and Multi-Photon Burst Size}

\noindent It is evident that the multi-photon burst increases the distance of stable transmission for all of the curves that we found. This is to be noted that this result can be generalized for same protocol at different multiphoton burst size. This can be done by using the following steps. To find this relation, we have to first find the \textit{turning} point, i.e., the point where the derivative of the curve becomes negative.
\begin{figure}[h!]
    \centering
    \includegraphics[width=0.48\textwidth]{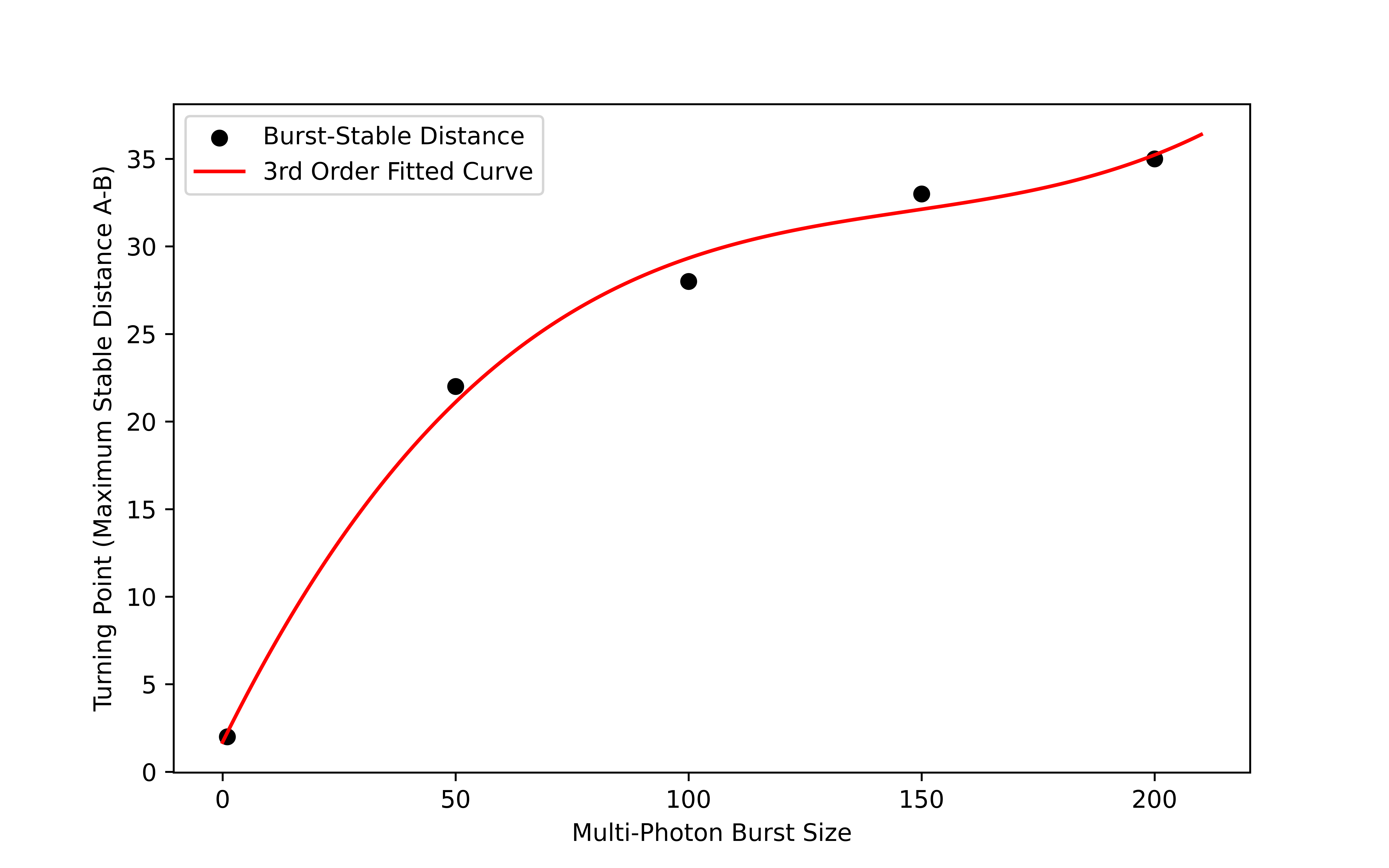}
    \caption{Relationship between the Maximum Stable Distance (A-B) and the size of multi-photon burst used.}
    \label{turning}
\end{figure}

\noindent From Fig.(\ref{turning}), we can see that a $3^{\text{rd}}$ order polynomial fit describes the relationship very well. Therefore, we fit a third-order polynomial to the data points and find the characteristic equation for the curve. The equation of the third-order polynomial curve was found as shown in equation(\ref{3rd-poly}).

\begin{equation}
    y = 0.000008x^3-0.003388x^2+0.538443x+1.693613
    \label{3rd-poly}
\end{equation}

\begin{figure}[h!]
    \centering
    \includegraphics[width=0.48\textwidth]{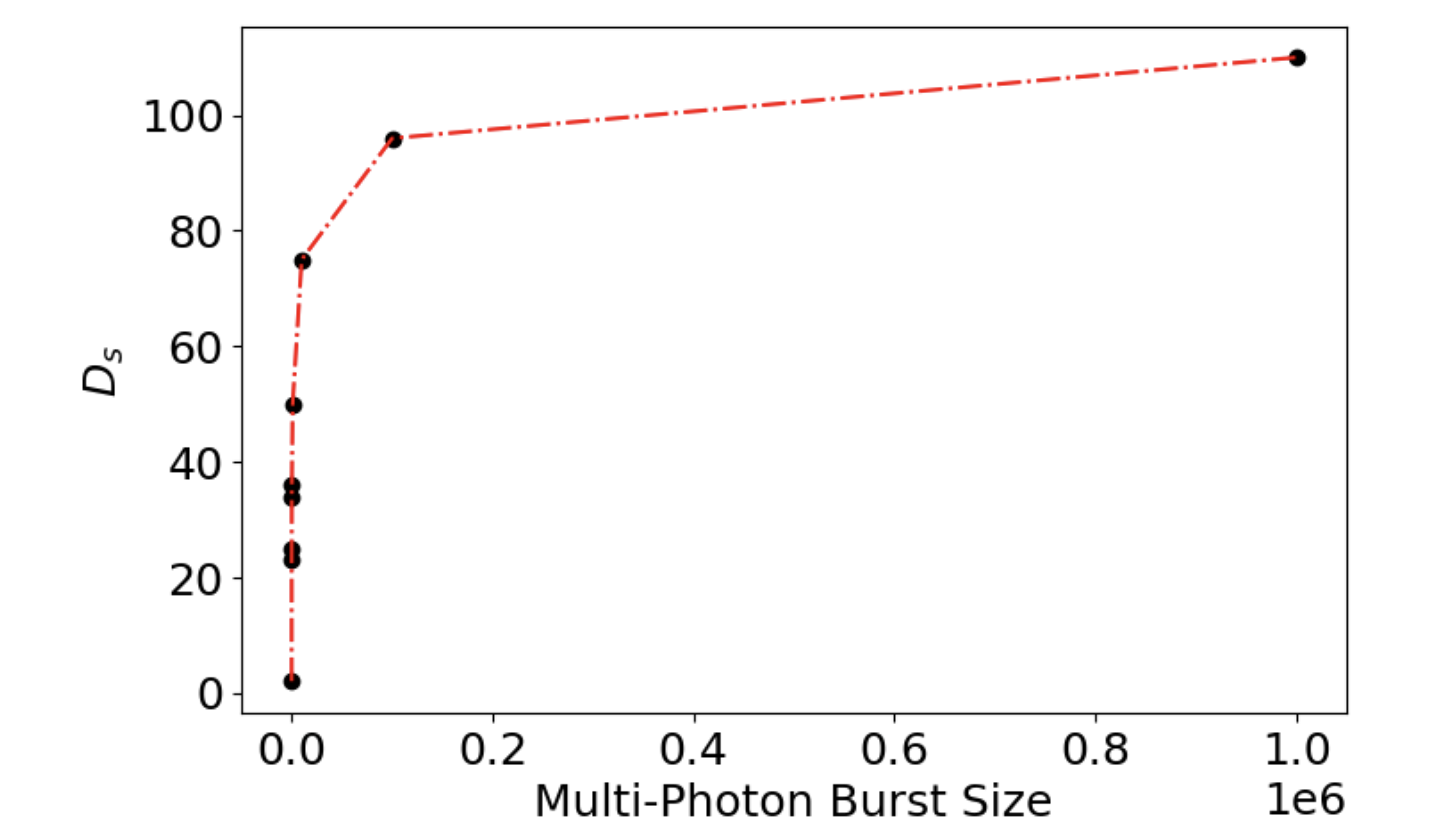}
    \caption{Relationship between the maximum stable distance (between Alice and Bob) and the size of multi-photon burst used consisting of burst sizes up to $10^6$ qubits. {The dashed lines is just for visualization purposes, and to study the data-trend more effectively, we do curve fitting in $\log$ scale in next figure.} }
    \label{fig:million-burst}
\end{figure}

\noindent From equation(\ref{3rd-poly}), we can see that multi-photon burst size follows a polynomial relationship with the distance of stable transmissions. However, the burst sizes used in this calculation are of very small order. To get a general picture and a more practical burst-size relation, we increase the burst size up to a million qubits at once. Fig.(\ref{fig:million-burst}) shows the curve with stable transmission distance and multiphoton burst size for bursts of up to one million qubits at once.

\begin{figure}[h!]
    \centering
    \includegraphics[width=0.48\textwidth]{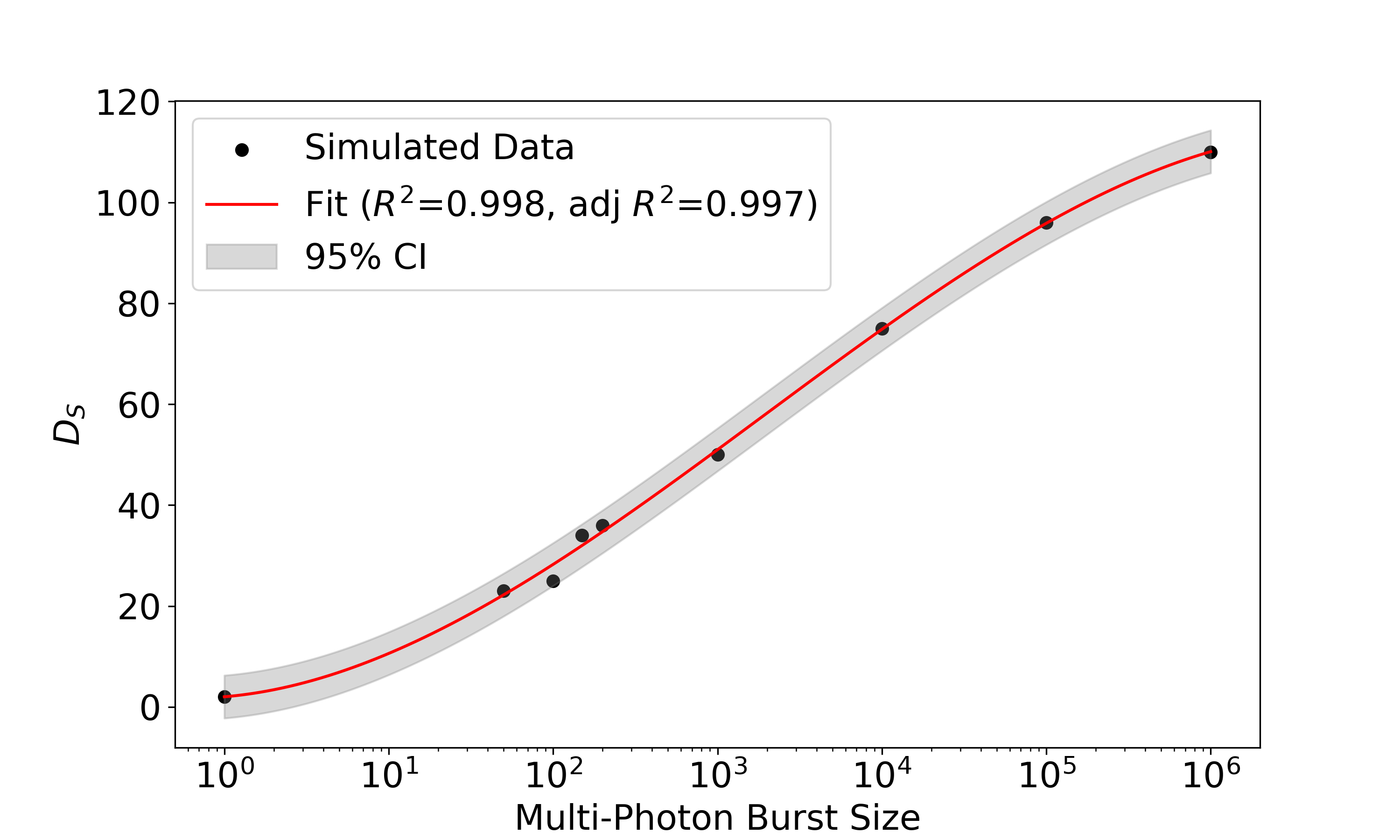}
    \caption{Relationship between the Maximum Stable Distance (A-B) and the size of multi-photon burst used consisting of burst sizes up to $10^6$ qubits (x-axis is in $\log$ scale). {\textit{CI} refers to the confidence index of the curve fitting.}}
    \label{fig:log-million}
\end{figure}

Fig.(\ref{fig:million-burst}) gives an insight into a generic relationship between the higher-order bursts and the distance of stable transmission. We set a curve between $\log(\text{burst\_values})$ and the distance of stable transmission. Fig.(\ref{fig:log-million}) gives us a better picture of a possible relationship between the main two quantities of interest of this study, that is, the distance of stable transmission and the size of multiphoton burst used. 

From Fig.(\ref{fig:log-million}), we found that the two quantities of interest share a logarithmic-$3^\text{rd}$ order polynomial relationship as described by equation(\ref{eq:relation-log}). {The model explains over $99\%$ of the variance ($R^2 = 0.998$) and remains high after adjusting for its four parameters (adjusted $R^2 = 0.997$), showing that a cubic fit in $\log(\mathrm{burst})$ works well. Furthermore, a confidence band of $95\%$ also shows very less uncertainty in the fit.}

\begin{equation}
D_s = -0.054x^3+1.228x^2+1.1878x+2.0178,
\label{eq:relation-log}
\end{equation}

where $x = \log(\text{burst\_size})$. Therefore, based on the above, we can define a generic relationship for a 3-stage protocol over a line topology between the distance of stable transmission (denoted by $D_s$), and the size of multi-photon burst used as described in equation(\ref{eq:generic}).

\begin{equation}
    D_s  = -\phi \log^3(b) + \beta \log^2(b) + \gamma \log (b) + \delta,
    \label{eq:generic}
\end{equation}

where, $\phi, \beta, \gamma, \text{ and } \delta$ are the curve fitting parameters and $b$ is the size of the multi-photon burst used. This gives us insight into a rather counter-intuitive relationship occurring between the two quantities. For line topology with $B=0.85$ and $D=0.02$, and $\alpha=0.15$, equation(\ref{eq:relation-log}) can be used to determine the maximum distance of stable transmission for a given value of the multi-photon burst.

\section{Conclusions and Future Work}
\label{Sec:Conclusion}

Quantum networks are increasingly becoming viable, as evidenced by empirical studies from initiatives such as the DARPA, European, and Tokyo quantum networks. These studies have addressed issues such as the distance between nodes and the stabilization of transmissions through the use of trusted nodes and optical switches. As indicated in equation(\ref{transprob}), the probability of successful transmission is dependent on the attenuation coefficient and the distance between the nodes, with different optical materials exhibiting variable attenuation constants. This requires exploring diverse topologies to optimize node placement and attenuation impacts. Our study evaluates the performance of three prominent QKD protocols—Decoy-state, 3-stage, and E91—across various network topologies.

Our findings indicate that the grid topology outperforms simpler configurations by leveraging multiple paths and trusted nodes, enhancing the key pool and robustness across the three explored QKD protocols. Notably, when we analyzed the torus topology with the $3$-stage and E$91$ protocols, it yielded significantly higher key rates than any other topology. Additionally, we derived a mathematical model to quantify key-rate variations with increasing distances between Alice and Bob over the line topology. We also formulated an equation for the maximum feasible distance for stable transmission in a line topology with three nodes, laying foundational insights for practical quantum network designs. This advancement underscores the importance of tailoring the multi-photon burst size to the node separation, optimizing communication efficacy in practical quantum networks. {Apart from these, the multiphoton approach studied in this work can be used in multi-carrier continuous‐variable QKD (CVQKD) systems, where coherent states are multiplexed into Gaussian subcarrier channels to increase aggregate key rates and resilience against channel noise}\cite{gyongyosi2020multicarrier}. {All of the results presented in this work define future frameworks for a more scalable and reliable network, laying foundation for the development of Quantum Internet}\cite{burr2022quantum}. 

As future work, we aim to establish a relationship between multi-photon burst sizes and maximum stable transmission distances across various topologies, enhancing our understanding of multi-photon device behaviors. There is also a need to address practical challenges associated with current QKD protocols, which often rely on heuristics and may not provide optimal solutions, thereby limiting their practicality for larger network development \cite{zeng2023entanglement}. Moreover, in the NISQ era, network engineering must account for Johnson noise and other real-world phenomena, necessitating the development of advanced error-correction methods to improve quantum router performance in noisy environments \cite{shi2023quantum}. A proposed solution to these problems in recent times has been quantum augmented networks. In this approach, several quantum components (such as quantum teleportation\cite{parakh2022quantum}, QKD, quantum error correction, etc.) are strategically integrated in current classical networks to enable large-scale quantum communication \cite{jha2025towards, jha2024ml}.

In summary, this study not only highlighted the advantages of multi-photon QKD protocols across different topologies but also established an empirical relationship between burst size and stable transmission distances for the 3-stage protocol. This protocol demonstrated enhanced capabilities for transmitting higher-order qubit bursts, yielding more stable and higher key rates. While the empirical relationship was specifically developed for the 3-stage protocol over the line topology, the insights gained will guide the selection of suitable protocols for crafting more robust quantum networks. 

\section*{Acknowledgment}
This work is sponsored by NSF award \#2324924 and \#2324925.

\printbibliography

\begin{IEEEbiography}[{\includegraphics[width=1in,height=1.25in,clip,keepaspectratio]{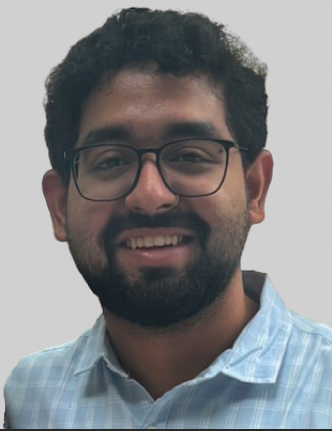}}]{Nitin Jha} 
received his BSc. (Hons) in Physics from Ashoka University in 2023. He is pursuing his Ph.D. from Kennesaw State University (KSU). His broad research interests lie in quantum networks, secure quantum communication, and network security. He is looking at application of artificial intelligence techniques to improve the efficiency of quantum networks and building the foundations of quantum-classical hybrid networks (also known as quantum augmented networks). Nitin's research has won multiple accolades at KSU, appeared in top venues, and continues to be a part of the strategic initiative at the University.
\end{IEEEbiography}

\begin{IEEEbiography}[{\includegraphics[width=1in,height=1.25in,clip,keepaspectratio]{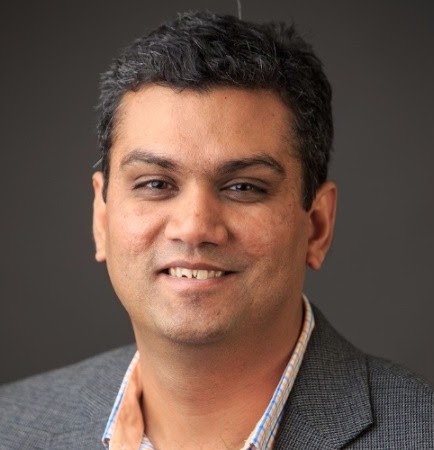}}]{Abhishek Parakh} 
is a Professor of Computer Science and the Director of the Computer Science Ph.D. Program at Kennesaw State University. He received his Ph.D. in Computer Science from Oklahoma State University in 2011. He has held various academic positions, including Director of NebraskaCYBER and the Mutual of Omaha Distinguished Chair of Information Science and Technology. His research interests include quantum cryptography, cybersecurity, and trustworthy computing. He has been the principal investigator on several federally funded projects (NSF, DOD, NASA, NSA, DOS) related to quantum networks, quantum computing, post-quantum cryptography, AI driven educational technology, and cybersecurity education.
\end{IEEEbiography}

\begin{IEEEbiography}[{\includegraphics[width=1in,height=1.25in,clip,keepaspectratio]{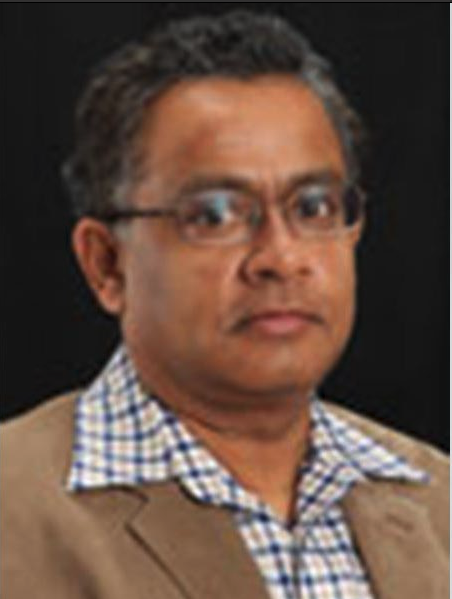}}]{Mahadevan Subramaniam.}  
is the Charles W. and Margre H. Durham Distinguished Professor and Chair of the Computer Science Department at the University of Nebraska Omaha (UNO). He earned his B.S. in Computer Science from BITS Pilani in 1986 and his M.S. and Ph.D. in Computer Science from SUNY Albany in 1997. His research focuses on formal methods for software engineering, particularly model checking, SMT solvers, and theorem provers, with applications in software evolution and automated repair strategies.
\end{IEEEbiography}

\end{document}